\documentclass[10pt]{article}
\pdfoutput=1
\usepackage{amsmath}
\usepackage{amssymb}
\usepackage{mathtools}
\usepackage{mathrsfs}
\usepackage{graphicx}
\usepackage{caption}
\usepackage{subcaption}
\usepackage{pdflscape}
\usepackage{pdfpages}
\usepackage{enumerate}
\usepackage{array}
\usepackage{makecell}
\usepackage{ctable}
\usepackage{morefloats}
\usepackage{setspace}
\usepackage{authblk}
\usepackage[]{appendix}
\usepackage{cite}
\usepackage{tabu}
\usepackage{geometry}
\usepackage[parfill]{parskip}
\usepackage[hidelinks]{hyperref}
\usepackage{titlesec}

\titleformat*{\section}{\large\bfseries}
\titleformat*{\subsection}{\normalsize\bfseries}
\titleformat*{\subsubsection}{\small\bfseries}

\begin{document}

\title{A Heuristic Reference Recursive Recipe for the Menacing Problem of Adaptively Tuning the Kalman Filter Statistics.\\ Part-1. Formulation and Simulation Studies.}
\author{M. R. Ananthasayanam$^1$ \footnote{sayanam2005@yahoo.co.in (corresponding author)} , Shyam Mohan M$^2$ \footnote{shyammoh.2014@iitkalumni.org} , Naren Naik$^3$ \footnote{nnaik@iitk.ac.in} , R. M. O. Gemson$^4$ \footnote{mogratnam@rediffmail.com}\\
{\normalsize $^1$Formerly Professor, Department of Aerospace Engineering, IISc, Banglore}\\
{\normalsize $^2$Formerly Post Graduate Student, IIT Kanpur, India}\\
{\normalsize $^3$Department of Electrical Engineering, IIT Kanpur, India}\\
{\normalsize $^4$Formerly Additional General Manager, HAL, Bangalore, India}}
\date{}
\maketitle

\textbf{Abstract.} Since the innovation of the ubiquitous Kalman filter more than five decades back it is well known that to obtain the best possible estimates the tuning of its statistics $\mathbf{X_0}$, $\mathbf{P_0}$, $\Theta$, \textbf{R} and \textbf{Q} namely initial state and covariance, unknown parameters, and the measurement and state noise covariances is very crucial. The earlier tweaking and other systematic approaches are reviewed but none has reached a simple and easily implementable approach for any application. The present reference recursive recipe based on multiple filter passes through the data leads to a converged `statistical equilibrium' solution. It utilizes the pre, post, and smoothed state estimates and their corresponding measurements and the actual measurements as well as their covariances to balance the state and measurement equations and form generalized cost functions. The filter covariance at the end of each pass is heuristically scaled up by the number of data points and further trimmed to provide the $\mathbf{P_0}$ for subsequent passes. A simultaneous and proper choice for \textbf{Q} and \textbf{R} based on the filter sample statistics and certain other covariances leads to a stable filter operation providing the results after few iterations. When only \textbf{R} is present in the data by minimizing the `innovation' cost function \textbf{J} using the non filter based Newton Raphson optimization results served as an anchor for matching and tuning the filter statistics. When both \textbf{R} and \textbf{Q} are present in the data the consistency between the injected noise sequences and their statistics provided a simple route and confidence in the present approach. A typical simulation study of a spring, mass, damper system with a weak non linear spring constant shows the present approach out performs earlier techniques. The Part-2 of the paper further consolidates the present approach based on an analysis of real airplane flight test data.


\section{Introduction}

Kalman proposed the solution for the linear filtering problem in his famous 1960 paper which is known as Kalman filter(KF). It is not well known to many that the enthusiasm which followed soon after Kalman introduced his filter was damped since the statistics of the process (\textbf{Q}) and measurement noise (\textbf{R}) had to be provided to design and implement the filter. Gauss had a relatively ideal situation with a good system model and only the measurements had noise and thus with his least squares approach he could get an estimate and a qualitative measure for the uncertainty. Kalman when he proposed the filter dealt with only state estimation. Presently the scale and magnitude of many difficult and interesting problems that estimation theory (ET) is handling could not have been comprehended by Gauss or Kalman. In many present day applications one does not even know quite well the structure of the state and measurement equations as well as the parameters in them and the statistical characteristics of the state and measurement noise. It is possible to add the unknown initial conditions of the state as well. One can summarize that almost nothing is known but everything has to be determined or estimated from the measurements alone! This means the connecting relationship between the state and the measurement has to be found out. This has to be from previous knowledge or intuition such that it is meaningful, reasonable, acceptable, or useful. Even this has to be achieved only by the internal consistency of the various quantities and/or the variables that occur at different times during the filter operation through the data. Further the systems are large and the best possible optimal estimation has to be worked out in real time. However due to the enormous computing power that is presently available it has been possible to handle the above situations. The other unknowns include system parameters ($\Theta$), the initial state $\mathbf{X_0}$ and its covariance $\mathbf{P_0}$. Tuning is done manually even today since the adaptive filtering method had not matured to a routine procedure or recipe. This is in spite of its applications in many fields of science and engineering including among many in airplane flight test data analysis (Klein and Morelli 2006), target tracking  (Bar-Shalom et al. 2001), evolution of the space debris scenario (Ananthasayanam et al. 2006), fusion of GPS and INS data (Grewal et al. 2007), study of the tectonic plate movements (Kleusberg and Teunissen 1996), high energy physics (Fruhwirth et al. 2000), agriculture, biology and medicine (Federer and Murthy 1998), dendroclimatology (Visser and Molenaar 1988), finance (Wells 1996), source separation problem in telecommunications, bio medicine, audio, speech, and  in particular astrophysics (Costagli and Kuruoglu 2007), and atmospheric data assimilation for weather prediction (Evensen 2009).

Instead of tweaking, an adaptive and systematic estimation of $\mathbf{X_0}$, $\mathbf{P_0}$, $\Theta$, \textbf{R} and \textbf{Q} is known as the problem of tuning the Kalman filter. In the present work, a reference recursive recipe (RRR) for tuning the Kalman filter is proposed. A simultaneous and proper choice for \textbf{Q} and \textbf{R} based on the filter sample statistics and certain other covariances leads to a stable filter operation providing the results after few iterations. We have also suggested an alternative statistic for the estimation of \textbf{Q} based upon the difference between the stochastic and dynamic trajectories (see Section-\ref{DSDT}). The RRR attains statistical equilibrium after few filter iterations (without any direct optimization) over the data together with internal consistency checks through different cost functions. The cost functions (see Section -\ref{RRR}) act as a pointer for the user to decide on the convergence of the filter which can otherwise be deceptive. Another important factor which is generally ignored is the Cramer Rao Bound (CRB) of the unknown parameters. This paper also discusses the importance of tuning the $\mathbf{P_0}$ and a simple heuristic scaling up method (see Section-\ref{choiceP0}) is proposed for achieving the CRBs.

The layout of the paper (Part-1) is as follows. The problem is set up in an extended Kalman filter (EKF) framework in Section-\ref{EKF}. The filter tuning and its importance is discussed in Section-\ref{Tuning}. A review of the earlier adaptive tuning methods and the present approach is discussed in Section-\ref{review} and \ref{RRR} respectively. The Section-\ref{simulation} discusses the results of the proposed reference recursive recipe (RRR) applied to a simulated spring mass and damper system with weak nonlinear spring constant. The concluding remarks are made in Section-\ref{conclusions}. The Part-2 of the paper shows the efficacy of the proposed RRR on more involved real airplane flight test data.

\section{Extended Kalman Filter Equations}
\label{EKF}
We consider an Extended Kalman Filter (EKF) formulation as it provides the best scenario since other filter formulations contain the effect of approximations, discretization and other features. Consider the following well known discrete time nonlinear filtering problem given by
\begin{align}
x_k&=f(x_{k-1},\Theta,u_{k-1})+w_k\\
Z_k&=h(x_{k},\Theta)+v_k \text{ , k = 1, 2,\ldots, N}
\end{align}
where `x' is the state vector of size $n\times 1$, `u' is the control input and `Z' is the measurement vector of size $m\times 1$. The `f' and `h' are non linear functions of state and measurement equations respectively. The process noise, $w_k\sim \mathbf{\mathcal{N}}$ ( 0, \textbf{Q}) and the measurement noise, $v_k\sim\mathbf{\mathcal{N}}$ ( 0, \textbf{R} ) are assumed to be zero mean additive White Gaussian Noise (WGN). The Normal or Gaussian distribution is represented by `$\mathbf{\mathcal{N}}$' and its assumed that,
\begin{align*}
E\left[w_kw_j^T\right]&=\textbf{Q }\delta(k-j) \text{ with } E\left[w_k\right]=0   \\
E\left[v_kv_j^T\right]&=\textbf{R }\delta(k-j) \text{ with } E\left[v_k\right]=0  \\
E\left[w_kv_j^T\right]&=0 \text{ $\forall$ j, k = 1, 2,\ldots, N }
\end{align*}
where  E$\left[\text{ }\right]$ is the expectation operator, $\delta$ is Kronecker delta function defined as\\
\[ \delta(k-j) = \left\{ \begin{array}{ll}
        0 & \mbox{if $k \neq j$};\\
        1 & \mbox{if $k = j$}.\end{array} \right. \]
In the EKF formulation the parameter vector `$\Theta$' of size $p \times 1$ is augmented as additional states leading to,
\begin{align*}
\begin{bmatrix}x_k \\ \Theta_k\end{bmatrix}=&
\begin{bmatrix}
f(x_{k-1},\Theta_{k-1},u_{k-1})\\
\Theta_{k-1}
\end{bmatrix}+
\begin{bmatrix}
w_k\\0
\end{bmatrix}
%X_k=&f(X_{k-1})+w_k
\end{align*}
The non linear filtering problem is now redefined as
\begin{align}
\label{e14} X_k&=f(X_{k-1})+w_k\\
\label{e15} Z_k&=h(X_{k})+v_k\text{ , k = 1, 2,\ldots, N}
\end{align}
where `X' and `w' are respectively the augmented state and process noise vector is of size $(n+p)\times 1$ and thus $w_k\sim \mathbf{\mathcal{N}} \left( 0, \begin{bmatrix}\textbf{Q} &0\\0&0\end{bmatrix} \right)$. The control input `u' and the `hat' symbol for estimates are not shown for brevity. A solution to the above problem is summarised in Brown and Hwang (2012),
\begin{table}[h!]
\begin{center}
%\begin{footnotesize}
\begin{tabular}{  l  l }
%\hline
\textbf{Initialisation :} & $\mathbf{X_0}=E\left[X_{t0}\right]$, $\mathbf{P_0}=E\left[(\mathbf{X_0}-X_{t0})(\mathbf{X_0}-X_{t0})^T\right]$ \\[10pt]
\textbf{Prediction step :} & $X_{k|k-1}=f(X_{k-1|k-1})$, ${P}_{k|k-1}=F_{k-1}{P}_{k-1|k-1}F^T_{k-1}+\textbf{Q}$ \\[10pt]
\textbf{Update step :} & $K_k={P}_{k|k-1}H^T_{k}(H_{k}P_{k|k-1}H^T_{k}+\textbf{R})^{-1}$ \\
& $X_{k|k}=X_{k|k-1}+K_k(Z_{k}-h(X_{k|k-1}))$, ${P}_{k|k}=(I-K_kH_{k}){P}_{k|k-1}$\\[12pt]
\multicolumn{2}{l}{where all the symbols have their usual meaning and }
\end{tabular}
%\end{footnotesize}
\end{center}
\end{table}
\begin{align*}
&\text{True initial state} &:& X_{t0} \\
&\text{Initial state estimate} &:& X_{0|0}=\mathbf{X_0} \\
&\text{Initial state covariance matrix} &:& P_{0|0}= \mathbf{P_0}\\
&\text{State Jacobian matrix}&:&F_{k-1}= \left[\frac{\partial{f}}{\partial{X}}\right]_{X=X_{k-1|k-1}}\\
&\text{Measurement Jacobian matrix}&:&H_{k}= \left[\frac{\partial{h}}{\partial{X}}\right]_{X=X_{k|k-1}}
\end{align*}

The above steps that statistically combine two estimates at any given time point, one from state $X_{k|k-1}$ and the other from measurement $Z_k$ equation are formal if only their uncertainties denoted by their covariances are given. The states can be estimated  given the initial $\mathbf{X_0}$ and $\mathbf{P_0}$ over a time span along with the process noise input with covariance \textbf{Q} and updated with measurements with noise covariance \textbf{R}. In order to match or minimize the difference $\nu_k =Z_k - h(X_{k|k-1})$ called the innovation in some best possible sense the estimates from the states and measurements over a length of data a well known criterion is the Method of Maximum Likelihood Estimation (MMLE). The innovation follows a white Gaussian distribution (Kailath 1970) which is operationally equivalent to minimizing the cost function
\begin{align*}
\mathbf{J}=&\frac{1}{N}\sum \mathbf{\nu_k}(H_kP_{k|k-1}H_k^T+\textbf{R})^{-1}\mathbf{\nu_k}^T\\
=&\mathbf{J}(\mathbf{X_0,P_0,Q,R},\Theta)\\
=&\mathbf{J}(\mathbf{X_0},\Theta,\mathbf{K}(\text{traded for } \mathbf{P_0, Q, R}))
\end{align*}
based on the summation over all the N measurements and thus solving for either $\mathbf{X_0, P_0, Q,}$\textbf{R}, $\Theta$ or solving for $\mathbf{X_0}$, $\Theta$, K as the case may be. When \textbf{Q} = 0, the MMLE is called as the output error method with the Kalman gain matrix being zero. In the usual Kalman filter implementation generally one does not solve for the statistics $\mathbf{P_0}$, \textbf{Q} and \textbf{R} but tweak manually to obtain acceptable values. The numerical effort of minimizing \textbf{J} has to appear in the estimation of the filter statistics. The Kalman filter is not a panacea to obtain better results when compared to simpler techniques of data analysis. The accuracy of the results using Kalman filter depends on its design based on the choice of $\mathbf{X_0}$, $\mathbf{P_0}$, $\Theta$, \textbf{R} and \textbf{Q}. If the above values are not chosen properly then the filter results can be inferior to those from simpler techniques.

There are five steps in the Kalman filter, namely state and covariance propagation with time, Kalman gain calculation and the state and covariance updates by incorporating the measurement. In the filter statistics approach all the five steps have to be gone through. The state propagation and update refer to the sample while the covariance propagation, update, and the Kalman gain refer to the ensemble characteristics.

In the present work we introduce a generalized cost function by an expansion of the usual `innovation' to other quantities generated by the Kalman filter such as the prior and posterior state respectively before and after the measurement is assimilated, the smoothed state after all the measurements are processed. We demonstrate that it is such an approach that decisively indicates the best possible solution in the simulated data analysis and more so in the analysis of real flight test data.

\section{Tuning of the Kalman Filter Statistics}
\label{Tuning}
Some fundamental differences exist between the cost function handled by classical optimization problems and the Kalman filter. The former deals with a static cost function, with a fixed model and a finite amount of data, with deterministic unknowns. In contrast the Kalman filter has to grapple with a dynamical time varying cost function due to possibly time varying state and measurement model structure and their parameters, with increasing number of measurements with the unknowns being both deterministic and the statistics of probabilistic variables. Obviously the effort to be put in minimizing \textbf{J} cannot be swept under the rug! Generally one manually tweaks the statistics to reach acceptable results instead of tuning properly to get even better results. In spite of its immense applications for more than five decades in many problems of science and technology the filter tuning has not matured to a simple and easily implementable approach for any problem. Even a routine adaptive filtering technique to estimate a constant signal with measurement noise does not appear to exist!

The ghost of filter tuning chases every variant and formulation of the filter be it EKF, Unscented Kalman Filter (UKF), Particle filter (PF), or the Ensemble Kalman Filter (EnKF) or their combinations. If one desires to obtain near optimal estimates then the filter statistics have to be properly tuned. In the absence of proper filter tuning it is difficult to infer if the performance of the variants of Kalman filter are due to their formulation or filter tuning!

In the best spirit of the estimation theory and in particular the recursive Kalman filter approach even if $\mathbf{X_0}$, $\mathbf{P_0}$, $\Theta$, \textbf{R} and \textbf{Q} are unknown or inaccurately known the filter should still have the ability to estimate all the above from the measurements Z and perhaps commencing not too far from their proper values. The filter results should prove to be self consistent and estimate all the unknowns. Generally one tunes the filter statistics off line using simulated data and later use it to process real data on line or even in real time.

The filter tuning varies from ad hoc, through heuristic to rigorous methods. Generally the tuning is manual or with adhoc quick fix solutions such as limiting \textbf{P} from going to zero, or add \textbf{Q} to increase \textbf{P} before calculating the gain, multiply \textbf{P} by a factor to limit \textbf{K} all have obviously limitations in handling involved problems or scenarios. In the fading memory filter the \textbf{R} and \textbf{Q} estimates are based on the weighting between the current data and previous estimate. All the above introduce additional parameters to be adjusted that varies for every problem. The rigorous approach could be hard and time consuming to the extent of solving the whole problem. It is the middle path of heuristic approaches which is quite appealing and followed in the present work. The adaptive approach or filter tuning tries to obtain the filter statistics $\mathbf{X_0}$, $\mathbf{P_0}$, $\Theta$, \textbf{R} and \textbf{Q} by using the various quantities arising in the filter while operating over the measurement data.

\subsection{Qualitative Features of the Filter Statistics}

Should the $\mathbf{P_0}$ $\equiv$ \textbf{Q} $\equiv$ 0 then the filter will learn nothing from the measurements and ignore it. The \textbf{R} is fairly objective and can be determined from the measured data. The $\mathbf{P_0}$ is tricky and generally the off diagonal elements are set to zero and the diagonal elements are set to large values but however their relative values are crucial for an optimum filter operation. The $\mathbf{P_0}$ controls the handover from the initial transient to \textbf{Q} that control the  steady state filter behavior. The \textbf{Q} though considered notorious helps to inject uncertainty into the state equations to assist the filter to learn from the measurements. Even when only the measurement noise is present in the data, starting with some initial estimate $\mathbf{X_0}$  somewhat away from the true values, in order to assist the filter to learn from the measurements a non zero \textbf{Q }has to be injected into the state equations since the effect of $\mathbf{P_0}$ will fade away quickly. A large value of \textbf{Q} will lead to a short transient with large steady state uncertainty of the estimates and vice versa for small \textbf{Q}. In the present work since both $\mathbf{X_0}$ and $\mathbf{P_0}$ are simultaneously tuned for data without process noise the requirement for injecting \textbf{Q} does not arise.

It is only by \textbf{Q} an analyst can handle the unmodelled or unmodellable errors. The process noise can estimate, account or offset for some deficiency, inaccuracy, or error in the initial conditions, system and measurement model equations, control or external input, measurement noise statistics, the numerical state and covariance propagation or update operations and even account for numerical  errors. Though \textbf{Q} is considered notorious it is the life line of the Kalman filter to do good work. The \textbf{Q} is helpful to track systems whose dynamical equations are unknown. Some classic examples are the GPS receiver clocks, satellite, trajectory of aircraft, missiles and reentry objects. These are handled by using the kinematic relations between the position, velocity, acceleration, and even jerk (Mehrotra and Mahapatra 1997) all assumed (or even estimated by postulating stochastic models with floated parameters) to be driven by white Gaussian noise \textbf{Q} of suitable magnitude to enable the filter to track these systems. The process noise inhibits the onset of instability.

There could be different methods of implementing the filter for any practical situation. One could work with time varying full, diagonal, or constant matrices \textbf{Q} and \textbf{R}, or work with constant Kalman gain matrix, or with important Kalman gain matrix elements. The constant \textbf{Q} and \textbf{R} matrices approach is generally preferable and the constant gain approach is more easily implementable though not necessarily optimal.

\subsection{Some Simple Choices for Initial $\mathbf{X_0}$ for States and Parameters}

For both the states and parameters the choice of $\mathbf{X_0}$ appears to be relatively weak for the filter estimates but the choice of $\mathbf{P_0}$ is very important in particular for the parameter uncertainty represented by Cramer Rao Bound (CRB). Generally even with not quite a good tuning of the filter statistics the estimates could be near the optimum but lead to large variation in the uncertainty represented by the CRB. However it may be cautioned that for data with only measurement noise if a Newton Raphson (NR) technique is used to minimize the cost function the effect of the initial unknown state can affect the estimates but can be handled if this is also treated as an unknown. In the case of the filter by the process of smoothing the unknown initial state can be also estimated.

Since some of the states are generally measured either the first or the average of the first few measurements can be taken as the initial value $\mathbf{X_0}$ for the state. The parameters are used as augmented states in the EKF route. Since either some computational or experimental results are available, these can be set as initial $\mathbf{X_0}$ for the parameter values. At times if possible a least square parameter estimates based on balancing the governing differential equations using the appropriate measurements can also be used as start up values.

\subsection{Importance of Initial $\mathbf{P_0}$ for States and Parameters}

If $\mathbf{P_0}$ is set equal to zero (for a very confident choice for the initial estimates) then the filter ignores and learns nothing from the measurements. If $\mathbf{P_0}$ is extremely large (a pessimistic choice for the initial values), then the filter believes the measurements much more and provides very little weight or ignores the state model values leading to large fluctuations in the state and parameter estimates along with large final uncertainty. Thus there appears to be a proper choice for $\mathbf{P_0}$ which is neither zero nor infinity to provide proper estimates and uncertainty.

This $\mathbf{P_0}$ is one of the important tuning parameters as stressed by very few like Maybeck  (1979), Candy (1986), and Gemson (1991) but most people treat it casually. Usually one assumes a guess $\mathbf{P_0}$ which tends to become very low after some data points and in order to make the filter learn from the subsequent measurements introduce an additional \textbf{Q} by trial and error into the state equations. This finally leads to some estimates and uncertainties. The former usually may be close but the latter generally away from the correct values. The peculiar situation is one has introduced \textbf{Q} even when there is no model structure uncertainty. In the present recursive procedure a proper $\mathbf{P_0}$ without any \textbf{Q} is shown to be possible.

The important point is that the initial state and parameter $\mathbf{P_0}$ can affect the final covariance (P$_{N|N})$ from the filter operation. This can be crucial in certain state estimation problems such as impact point estimation and its uncertainty for target tracking. In such cases the results for the uncertainty from improper or inaccurate tuning can be deceptive. Even in parameter estimation problems the estimates and the assigned uncertainties can be important in the design of control systems.

\subsection{Choice of Initial \textbf{R} and \textbf{Q}}
Usually a good initial estimate for \textbf{R} can be obtained from the calibration of the measuring instrument and generally it is assumed to be constant over the data length. In the NR procedure of minimizing the cost function based on data without process noise even \textbf{R} can be treated as an unknown and estimated along with other unknown quantities. In principle the \textbf{Q} should reflect the uncertainty in the assumed state model or any `unmodellable' feature of the state or even unknown random state input. The \textbf{Q} along with the initial $\mathbf{P_0}$ plays a very important role in the filter operation without divergence. The value of \textbf{Q} should be small enough to retain the learning potential from the measurement but not large to make the filter estimates useless.

\subsection{Simultaneous Estimation of \textbf{R} and \textbf{Q}}

\par When the data contains the effect of both the measurement and process noise it becomes far more difficult and notorious for analysis. In general minimization of any cost function by treating \textbf{R} and \textbf{Q} as unknowns during which the simultaneous update and estimation of \textbf{R} and \textbf{Q} appears to be difficult as reported by many researchers in the field. For a given \textbf{R} changing the sequence of measurement noise makes the dynamics noisy meaning more blurred. In contrast the varying sequence of process noise makes the dynamical system wander randomly. The interesting point is the filter by tracking the drifted dynamical behaviour even with large \textbf{Q} estimates the parameters controlling the original dynamics of the system without the effect of \textbf{R} and \textbf{Q}. Since \textbf{R} and \textbf{Q} occur respectively in the measurement and state equations their effects are negatively correlated. The \textbf{Q} represents the average rate of change of the state, tracking ability by indirectly controlling the rate at which the old data are forgotten, and affects the estimation accuracy. The effect of \textbf{R} is opposite to that of \textbf{Q}. Thus during simultaneous recursive estimation if the statistics for estimating them are not properly chosen then \textbf{R} is over estimated and \textbf{Q} is under estimated and vice versa. This is just the reason due to which Gemson (1991) in his approach had to update \textbf{R} and \textbf{Q} alternately.

\subsection{Tuning Filter Statistics with only \textbf{R}}

\par In the spirit of recursive filtering operations we have sought to tune all the filter statistics by repeated iterative processing of the data without any direct optimization procedures. When only measurement noise is present in the data it is possible to get both the estimates and the CRB by minimizing the cost function \textbf{J} formed by the difference of the estimated and the actual measurement. The NR procedure carries out the above task very efficiently for simple to involved systems (Ananthasayanam et al. 2001). The results from the above NR procedure served as an anchor for tuning the filter parameters to get the closest possible estimates and the CRB from the present recursive procedure. It appears to be not feasible to reproduce the exact NR results from the filter for each and every data but sufficiently close.

\subsection{Tuning Filter Statistics with both \textbf{R} and \textbf{Q}}
However when process noise is also present in the data without solving the involved optimization problem we look for consistency based on a comparison between the injected and estimated sequences of \textbf{R} and \textbf{Q}. Further the additional cost functions proposed in our work based on balancing the state and measurement equations helped to obtain confidence in the results. When the filter operates through the data it generates prior, post, and smoothed state estimates and their covariances which help to generate candidate `statistic' to estimate both \textbf{R} and \textbf{Q}. From these one can also generate additional cost functions to see how best the state equations are balanced. Then all the above state estimates and covariances can be transformed into measurement space to be compared once again with the actual measurements and its covariance to generate candidate `statistic' to estimate \textbf{R}. These help to form more cost functions to see how best the measurement equations are balanced. It is necessary to choose from among the above `statistics' the proper combination of `statistic' for simultaneously estimating \textbf{R} and \textbf{Q}. Any improper combination does not lead to proper filter operation and even if it does leads to inappropriate results as is the case of some earlier approaches to solve the filter tuning problem. The subsequent sections after a review of earlier approaches propose the present method. This paper demonstrates the results using the simulated data of a spring, mass, and damper system with a weak non linear spring followed by real data in the second part of the paper.

\section{Review of Earlier Adaptive Kalman Filtering Approaches}
\label{review}
One brute force method for tuning filter statistics is to carry out an optimization exercise to solve for the statistics based on minimizing a suitable cost function over thousands of combinations of the unknowns. These would be too unwieldy requiring such a massive numerical exercise to be carried out for each and every new problem scenario. We should thus look for elegant approaches which help tune the filter with reasonable numerical effort. In particular in EKF if the unknown noise covariances are incorrectly specified biased estimates can arise (Ljung 1979, Ljungquist and Balchen 1994). Even when the system parameters are known, if an inaccurate description of the noise statistics are used the filter may give poor estimates, or even diverge. The following sections discuss the four broad adaptive filtering techniques.

\subsection{Bayesian Method}
Every update in a Kalman filter is obviously a Bayesian update. The work of Alspach  (1974) deals with a bank of autonomous Kalman filter run with a range of Kalman gains. Each one stores a running sum of the square of the residuals. Subsequently it is possible to obtain the estimates of the unknowns based on a weighted sum over the grid points of the gain. Hilborn and Lainiotis (1969) show that a Bayesian optimal adaptive estimation system converges to the average performance of an (unrealizable) optimal system operating with knowledge of the parameters which are unknown to a Bayesian adaptive/learning system.

\subsection{Maximum Likelihood (ML) and Expectation Maximization (EM)}

\par The ML and EM methods need the specification of the probability distribution (usually a Gaussian) followed by the difference between the measurement and the model prediction called innovation or prediction error. The maximum likelihood methods (Kashyap 1970, Bohlin 1976) maximize the likelihood function based on the innovations containing the unknown covariances \textbf{Q} and \textbf{R}. Usually time consuming gradient based numerical optimisation or other schemes that reprocess the data several times are required to estimate the unknown covariances.

\par In order to avoid nonlinear optimization in the ML approach Shumway and Stoffer (2000) proposed an iterative method using the expectation maximization (EM) technique. The EM consists of expectation and maximization steps. First the states are estimated using an initial guess of the unknown parameters based on a Kalman smoother. The unknown parameters are next estimated by ML method. This process of estimating the states using the Kalman smoother and optimizing the parameters using is repeated until parameter convergence.

\par Bavdekar et al. (2011) used both a direct optimization method and an extended EM method for nonlinear systems, based on the extended Kalman filter. Their first approach directly optimizes the likelihood function of the innovation sequence generated by the EKF using sequential quadratic programming. The second approach using the extended EM method, maximizes the likelihood function of the complete data set of the states and measurements in an iterative procedure.

\par Recently Zagrobelny and Rawlings (2014) similarly proposed a ML method to identify \textbf{Q} and \textbf{R} assuming the parameters are known. By writing the outputs in terms of the process and measurement noises, they form a normal distribution for the sequence of measurements. The variance of this distribution is a function of the unknown noise covariances, and the likelihood is optimized with respect to these covariances. Simulations are used to compare the ML method to several existing methods like the autocovariance least squares (ALS) method, an alternate ML method based on the innovations, and an EM method.

\subsection{Covariance Matching}

\par During the filter pass over the data a number of random variables such as pre, post, and smoothed states arise and these transformed to the measurements along with actual measurements many more statistics are available. Combining these one can form innovation, residue, smoothed residues and so on. As mentioned earlier among the five filter equations two refer to the sample and the other three refer to the ensemble characteristics. If the filter is performing well then the sample statistics formed by the above should be internally consistent with their ensemble properties also provided by the filter. These can be matched by running many Monte Carlo simulations. However taking sample statistics at various time instants during a filter pass one can form equations connecting the many random variables, their combinations and their covariances to estimate the unknown covariances. The different approaches of Myers and Tapley (1976), Gemson (1991), Mohamed and Schwarz (1999), Bavdekar et al. (2011), and the present approach are examples for the covariance matching. Different statistics are chosen either over a window or the full data length, after each pass for estimating \textbf{R} and \textbf{Q}. These estimates could be used in subsequent passes as is necessary. Myers and Tapley (1976) approach using the innovation (available before update) can at times make the \textbf{R} lose its positive definiteness. This is because \textbf{R} depends on the difference of two matrices which could become negative. In order to overcome such an eventuality they proposed a remedy of simply turning the estimate as positive! Mohamed and Schwarz (1999) have suggested a more stable statistic based on the residue (available after update) for estimating \textbf{R} which is the sum of two matrices. The \textbf{R} estimates are generally more accurate than \textbf{Q}. Simultaneous estimation of \textbf{R} and \textbf{Q }is not easy perhaps as they negatively affect the filter performance. These covariance matching techniques are preferred due to their simplicity and speed despite being suboptimal. Gemson (1991) and Gemson and Ananthasayanam (1998) showed the importance of $\mathbf{P_0}$ for parameter estimation. Similarly many statistics are available and are utilized for estimating \textbf{Q}.

\subsection{The Correlation Technique}

The innovation theorem of Kailath (1970) states that the innovation sequence is zero mean white Gaussian. Kailath further stated that if any gain other than the optimal is used then estimates will be suboptimal and the innovation mean will be non zero and the autocorrelation will not follow the Kronecker delta function. Also if the covariance of the innovations are not as expected they are indicative of the choice of any or all of the system matrices as well as the covariances are incorrect.

The correlation technique pioneered by Mehra (1970, 1972) and Carew and Belanger (1973) and Belanger (1974) is the earliest and highly cited method for determining the unknown covariances. Mehra's approach is based on the properties of the innovations process that must be white and Gaussian. Starting from an assumed value for the unknown \textbf{R} and \textbf{Q} an initial estimate for the steady state Kalman gain is obtained. This sequence is checked to see whether the particular Kalman gain implemented generates a statistically acceptable white noise sequence. However the Kalman gain can take correct value even when \textbf{R} and \textbf{Q} are far away from their true values. This is because different combinations of \textbf{R} and \textbf{Q} can lead to the same gain.

Later Neethling and Young (1974) noted large covariances from Mehra's approach due to some other deficiencies. The approach of Oussalah and De Schutter (2000) improves Mehra's (1970) and Carrew's and Bellanger's (1973) approaches by incorporating information about the quality of the innovation estimates leading to a weighted least squares methodology. The weights are determined using a Bhattacharyya distance criterion between the ideal probability and the distribution referring to the current first and second order statistics of the autocorrelation functions. The latter helps to generate a convergent sequence to the steady state filter, which after some manipulations allows one to determine the values of the noise statistics \textbf{Q} and \textbf{R}.

The latter work of Odelson et al. (2006) showed based on some counter examples the mathematical conditions regarding the system and measurement matrices are not sufficient and not necessary in Mehra's work (1970). Also they showed that the variance estimates of Mehra are larger and often negative as well that is unacceptable. They proposed a method called constrained Autocovariance Least Squares (ALS) method corrected the above and obtained a much smaller and better estimates and none negative as well.

\subsection{Other Approaches}

\par Valappil and Georgakis (2000) used the available information about the model parametric uncertainties and translate this information to the process noise covariance \textbf{Q}. They propose two methods called linearized and Monte Carlo approaches. They also assume that \textbf{R} is readily available from the measuring instrument characteristics. The Monte Carlo simulations are run with parameter values sampled from the assumed normal distribution, with means and covariances. Then the difference between the nominal and the random dynamical state trajectory over many simulations is taken to provide the process noise at any time instant to be used in the filter equations.

\par Some attempts have been made like Powell (2002) using the simplex method, Oshman and Shaviv  (2000) using the genetic algorithm, and controlled random search (Anilkumar 2000). However when the dimension, nonlinearlty, and the range of search space become large these could become computationally prohibitive and could lead to local minimum.

\par Manika Saha et al. (2014) felt that $\mathbf{X_0}$ and $\mathbf{P_0}$ are not critical since they affect only the initial transient filter performance and $\mathbf{P_0}$ has to be decided by designers since \textbf{P} changes as the filter operation proceeds reaching a steady state if only the system dynamics does not substantially change and a suitable choice of \textbf{R} can be easily made. Thus for a chosen suitable values for $\mathbf{X_0}$, $\mathbf{P_0}$, and \textbf{R} they identified the critical \textbf{Q} by forming two metrics based on the innovation covariance. These vary from zero to the number of measurements and vice versa as \textbf{Q} changes from zero to infinity they proposed that \textbf{Q} can be chosen around the intersection point of these two cost functions.

\par Lau and Lin (2011) also discuss the limitations of simulated annealing and particle optimization techniques for filter tuning.  One can summarize that deterministic or probabilistic optimization approaches do not appear to be efficient for solving the filter tuning problem. Hence we tried to see if a recursive filtering approach would work and fortuitously it had worked and will be demonstrated subsequently.

\section{Present Recursive Reference Recipe for Tuning Filter Statistics}
\label{RRR}
\par Fundamentally the Estimation Theory (ET) is an optimization problem. Hence a suitable cost function \textbf{J} has to be chosen. Essentially there are two elements in ET (i) Defining a cost function, (ii) Adopting a suitable algorithm to minimize the cost function. The Likelihood cost (\textbf{L}) for normally distributed error (e) (Bohn 2000) is given by
\begin{align}
\mathbf{L}(\Theta)=&\frac{1}{N}\sum_{k=1}^N \frac{1}{2}(e_k)^T\mathbf{A_k}^{-1}(e_k)+log(det(\mathbf{A_k}))
\end{align}
where \textbf{A} is the error covariance matrix and det(\textbf{A}) represent determinant of matrix \textbf{A}. It may be noted that the unknown parameters ($\Theta$) occurs implicitly and not explicitly in the cost function \textbf{L}. Since the filter provides many quantities it is possible to have many more terms in the cost function which is a function of errors in the initial state and parameter estimates, process noise driving the system and the measurement noise introduced by the measurement system, each weighted appropriately through suitable weighting covariance matrices. Thus the new cost function \textbf{J} in a weighted least square sense accounts for (i) A priori knowledge about the initial estimates, and (ii) Balancing the measurement equations. (iii) Balancing the system equations. One can call this as generalized MLE (\textbf{J}),
\begin{align}
\mathbf{J = J0+ J1 + J2 + J3 +J4 + J5 + J6 + J7 + J8}
\end{align}
whose terms are defined as
\begin{align*}
\mathbf{J0}=&\frac{1}{2}(\mathbf{X_0}-X_{t0})^T\mathbf{P_0}^{-1}(\mathbf{X_0}-X_{t0})\\
\mathbf{J1}=&\frac{1}{N}\sum_{k=1}^N (Z_k-h(X_{k|k-1}))^TS1_k^{-1}(Z_k-h(X_{k|k-1}))\\
\mathbf{J2}=&\frac{1}{N}\sum_{k=1}^N (Z_k-h(X_{k|k}))^TS2_k^{-1}(Z_k-h(X_{k|k}))\\
\mathbf{J3}=&\frac{1}{N}\sum_{k=1}^N (Z_k-h(X_{k|N}))^TS3_k^{-1}(Z_k-h(X_{k|N}))\\
\mathbf{J4}=&\frac{1}{N}\sum_{k=1}^N (Z_k-h(Xd_{k|N}))^T(Z_k-h(Xd_{k|N}))\\
\mathbf{J5}=&\frac{1}{N}\sum_{k=1}^N (Z_k-h(X_{k|k-1}))^TS1_k^{-1}(Z_k-h(X_{k|k-1}))+log(det(S1_k))\\
\mathbf{J6}=&\frac{1}{N}\sum_{k=1}^N   w1_{k|N}^TW1_k^{-1}  w1_{k|N}\\
\mathbf{J7}=&\frac{1}{N}\sum_{k=1}^N   w2_{k|N}^TW2_k^{-1}  w2_{k|N}\\
\mathbf{J8}=&\frac{1}{N}\sum_{k=1}^N   w3_{k|k}^TW3_k^{-1}  w3_{k|k}
\end{align*}
The `S' and `W' are functions of the second order moments given by Shyam et al. (2015)
\begin{align*}
S1_k=&H_{k}P_{k|k-1}H_{k}^T+\textbf{R}\\
S2_k=&-H_{k|k}P_{k|k}H_{k|k}^T+\textbf{R}\\
S3_k=&-H_{k|N}P_{k|k-1}H_{k|N}^T+\textbf{R}\\
W1_k=&-P_{k|N}-F_{k-1|N}P_{k-1|k-1}F_{k-1|N}^T+P_{k,k-1|N}F_{k-1|N}^T+P_{k,k-1|N}^TF_{k-1|N}+\textbf{Q}\\
W2_k=&-P_{k|N}-Fd_{k-1|N}P_{k-1|k-1}Fd_{k-1|N}^T+P_{k,k-1|N}Fd_{k-1|N}^T+P_{k,k-1|N}^TFd_{k-1|N}+\textbf{Q}\\
W3_k=&P_{k|k-1}-P_{k|k}
\end{align*}

If the initial states are known then \textbf{J0} is not necessary but if they are unknown, their estimate and covariance can be obtained respectively by the smoothed estimates $\mathbf{X_{0|N}}$ and  $\mathbf{P_{0|N}}$.  The cost \textbf{J1, J2} and \textbf{J3} are expected to tend towards the number of measurements (m). The cost \textbf{J6, J7} and \textbf{J8} defined for states with process noise are expected to tend towards the number of states (n). The cost \textbf{J4} is expected to tend towards the trace of \textbf{R} for \textbf{Q} = 0 case and \textbf{J5} is the negative log likelihood function.

\subsection{Choice of $\mathbf{X_0}$}
\label{RTS}
\par Commencing from an assumed reasonable initial choice for $\mathbf{X_0}$, $\mathbf{P_0}$, $\Theta$, \textbf{R} and \textbf{Q} the first filter pass through the data is made. Then a backward smoothing procedure is carried out. Rauch, Tung and Striebel (RTS, 1965) smoother was used as a standard smoothing procedure throughout the present work. The smoothing leads to the best possible state and parameter estimates as well as their covariances. It may be noted after smoothing the state estimates and their covariances change but not that of the parameters. We next describe how the above choices are updated for further filter passes through the data to eventually converge which denotes statistical equilibrium. If the exact value of $\mathbf{X_0}$ if not given it can be obtained by the smoothed estimates. In such probabilistic approaches any number of unknowns are only determined in a probabilistic sense and thus contain uncertainty due to the percolation of all the noise effects over all the unknowns.  The RTS smoothing equations for discrete time instants k = N-1, N-2,\ldots ,0 are given by
\begin{align}
\left.\begin{array}{r@{\;}l}
K_{k|N}=&{P}_{k|k}F_{k}{P}_{k+1|k}^{-1}\\
X_{k|N}=&{X}_{k|k}+K_{k|N}({X}_{k+1|N}-{X}_{k+1|k})\\
P_{k|N}=&{P}_{k|k}+K_{k|N}({P}_{k+1|N}-{P}_{k+1|k})K_{k|N}^T
\end{array}\right\}\label{P3}
\end{align}
where $K_{k|N},{X}_{k|N}$ and $P_{k|N}$ is the smoothed gain, smoothed state estimate and smoothed state covariance matrix respectively.

\subsection{Choice of $\mathbf{P_0}$}
\label{choiceP0}
\par The choice of $\mathbf{P_0}$ for the next filter pass is very tricky. If one were to take the smoothed initial state covariance ($P_{0|N}$) and use it as the $\mathbf{P_0}$ for the next pass then the final covariance keep on decreasing with further filter passes and eventually tend towards zero. We know that such a fact is unrealistic. In order to remedy the above behaviour the final covariance at the end of the pass was scaled up by the number of data points (N) and used at the beginning of the next pass. The only heuristic reasoning that can be provided from statistics is based on the fact that the mean from a sample has an uncertainty \textbf{P} that keeps decreasing with sample size as \textbf{P}/N where \textbf{P} is the population variance. Since in the filter steps the estimates and its update refer to the sample and the other covariance propagation, its update, and the calculation of the Kalman gain refer to the ensemble characteristics before every filter pass we carry out the above scale up method to obtain the $\mathbf{P_0}$ for the next filter pass. We also propagated backwards the final covariance using the estimated parameters, and it turned out that there was not much of a difference with the $\mathbf{P_0}$ that was used in the forward pass. The scaled up $\mathbf{P_0}$ (Shyam 2014) is given by
\begin{align}
\label{P1} \mathbf{P_0}=N\times \mathbf{P}_{N|N}
\end{align}

The usual recommendation (Mehra 1970) when the states are measured is to set $\mathbf{P_0}$ = \textbf{R}. Even by using the Inverse of Information Matrix (IIM) approach of Gemson (1991) obtained the same estimate for $\mathbf{P_0}$ as \textbf{R}. The IIM is given by
\begin{align}
\label{P2}  \mathbf{P_0}=\left[\frac{1}{N}\sum_{k=1}^{N} F_{k-1}^TH_k^TR^{-1}H_kF_{k-1} \right]^{-1}
\end{align}

Of course the above process does not end and we have to further  trim the above $\mathbf{P_0}$ to obtain the best possible CRB after some passes. The scaled up $\mathbf{P_0}$ is a full  matrix. Many changes such as using only the diagonal elements and many more variations were  tried out. Finally the reference $\mathbf{P_0}$ to obtain the proper CRB for the parameter estimates turns  out to have all the elements are zero (the covariance of all the states and their cross covariance with the parameters as well) except the diagonal elements  corresponding to the parameters. If all the elements of the parameter covariances were included and the state and its cross covariances set to zero, it did not make much of a difference in the final results.

\subsubsection{Probability Matching Prior Interpretation for $\mathbf{P_0}$}

If the deterministic NR optimization approach is considered to provide frequentist results then the Kalman filtering approach corresponds to the Bayesian route. It is useful to recall the immense amount of study for the choice of suitable prior distributions in the Bayesian approach for inference in statistics (Datta and Mukerjee 2005). The choice of appropriate probability distribution for $\mathbf{P_0}$ is the probability matching prior (PMP) that provides a bridge between the above approaches to provide comparable results.

Consider the simple case of a constant signal with noise. In the frequentist approach the calculation of the mean and standard deviation is direct. However in the Bayesian approach the above result is not reachable unless a proper $\mathbf{P_0}$ is chosen which is the probability matching prior and in our case the tuning. Further we have in addition the process noise input to be handled.

\subsection{Estimation of \textbf{R} and \textbf{Q} using EM Method}
\par The expression for \textbf{R} and \textbf{Q} using the Expectation Maximisation (EM) method extended to a non-linear system by Bavdekar et al. (2011) is given by,

\begin{align}
%\left.\begin{array}{r@{\;}l}
\label{max1} \textbf{R}&= \frac{1}{N}\sum_{k=1}^N E\left[v_kv_k^T|Z_N\right]\\
\label{max2} \textbf{Q}&= \frac{1}{N}\sum_{k=1}^N E\left[w_kw_k^T|Z_N\right]
%\end{array}\right\}
\end{align}

\subsubsection{Estimation of \textbf{R}}
\par Consider the measurement noise, $v_k=Z_k-h(X_k)$ which can be approximated using first order Taylor series expansion around the smoothed estimate, $X_{k|N}$ given by
\begin{align*}
&v_k \approx Z_k-h(X_{k|N})-H_{k|N} \tilde X_{k|N}
\end{align*}
where $H_{k|N}=\frac{\partial{h}}{\partial{X}}_{|X=X_{k|N}}$ and $\tilde X_{k|N}=X_k-X_{k|N}$
\begin{align}
\nonumber &v_kv_k^T=Z_kZ_k^T-Z_kh^T(X_{k|N})-Z_k\tilde X_{k|N}^T H_{k|N}^T-h(X_{k|N})Z_k^T+h(X_{k|N})h^T(X_{k|N})\\
\nonumber &+h(X_{k|N})\tilde X_{k|N}^T H_{k|N}^T-H_{k|N}\tilde X_{k|N}Z_k^T+H_{k|N}\tilde X_{k|N}h^T(X_{k|N})+H_{k|N}\tilde X_{k|N}\tilde X_{k|N}^T H_{k|N}^T
\end{align}
We know that
\begin{align*}
&E[\tilde X_{k|N}]=E[X_k-X_{k|N}]=0\\
&E[\tilde X_{k|N} \tilde X_{k|N}^T]=E[(X_k-X_{k|N})(X_k-X_{k|N})^T]=P_{k|N}
\end{align*}
Thus the conditional expectation is given by
\begin{align*}
E\left[v_kv_k^T|Z_N\right]=Z_kZ_k^T-Z_kh^T(X_{k|N})-h(X_{k|N})Z_k^T+h(X_{k|N})h^T(X_{k|N})+H_{k|N}P_{k|N} H_{k|N}^T
\end{align*}
Rearranging the above terms and using Eq-\ref{max1}, we get
\begin{align}
\textbf{R}=\frac{1}{N}\sum_{k=1}^{N}\left\lbrace(Z_k-h(X_{k|N}))(Z_k-h(X_{k|N}))^T+H_{k|N}P_{k|N} H_{k|N}^T\right\rbrace
\end{align}

\subsubsection{Estimation of \textbf{Q}}
\par Consider the process noise, $w_k=X_k-f(X_{k-1})$ which can be approximated using first order Taylor series expansion around the smoothed estimate, $X_{k-1|N}$ given by
\begin{align*}
w_k \approx X_k-f(X_{k-1|N})-F_{k-1|N} \tilde X_{k-1|N}
\end{align*}
where $F_{k-1|N}=\frac{\partial{f}}{\partial{X}}_{|X=X_{k-1|N}}$ and $\tilde X_{k-1|N}=X_{k-1}-X_{k-1|N}$.
\begin{align}
\nonumber w_kw_k^T&=X_kX_k^T-X_kf^T(X_{k-1|N})-f(X_{k-1|N})X_k^T+f(X_{k-1|N})f^T(X_{k-1|N})\\
\nonumber &-X_k\tilde X_{k-1|N}^T F_{k-1|N}^T+f(X_{k-1|N})\tilde X_{k-1|N}^TF_{k-1|N}^T-F_{k-1|N}\tilde X_{k-1|N}X_k^T\\
\label{e11} &+F_{k-1|N}\tilde X_{k-1|N}f(X_{k-1|N})^T+F_{k-1|N}\tilde X_{k-1|N}\tilde X_{k-1|N}^T F_{k-1|N}^T
\end{align}
The following results are used in calculating the conditional expectation in Eq-\ref{max2}
\begin{align}
\left.\begin{array}{r@{\;}l}
E[X_kX_k^T|Z_N]&=X_{k|N}X_{k|N}^T+P_{k|N}\\
E[X_kX_{k-1}^T|Z_N]&=X_{k|N}X_{k-1|N}^T+P_{k,k-1|N}
\end{array}\right\}\label{r1}
\end{align}
The lag-one covariance, $P_{k,k-1|N}$ for k = N-1, N-2,\ldots, 1 is given by
\begin{align*}
& P_{k,k-1|N}=E[(X_k-X_{k|N})(X_{k-1}-X_{k-1|N})^T]\\
& P_{N,N-1|N}=(I-K_NH_N)F_{N-1}P_{N-1|N-1}\\
& P_{k,k-1|N}=P_{k|k}K_{k-1|N}^T+K_{k|N}(P_{k+1,k|N}-F_{k}P_{k|k})K_{k-1|N}^T
\end{align*}
where $K_{k|N}$ is the smoothed gain at discrete time `k' obtained from the RTS smoothing algorithm-\ref{RTS}. Using Eq-\ref{e11} and Eq-\ref{r1} we get the conditional expectation as
\begin{align*}
&E\left[w_kw_k^T|Z_N\right]=X_{k|N}X_{k|N}^T+P_{k|N}+X_{k|N}f^T(X_{k-1|N})-f(X_{k-1|N})X_{k|N}^T\\
\nonumber &+f(X_{k-1|N})f^T(X_{k-1|N})-P_{k,k-1|N}F_{k-1|N}^T-P_{k,k-1|N}^TF_{k-1|N}+F_{k-1|N}P_{k-1|k-1}F_{k-1|N}^T
\end{align*}
Rearranging the above terms and using Eq-\ref{max2}, we get
\begin{footnotesize}
\begin{align}
\textbf{Q}=\frac{1}{N}\sum_{k=1}^N &\{  w1_{k|N}   w1_{k|N}^T+P_{k|N}+F_{k-1|N}P_{k-1|N}F_{k-1|N}^T-P_{k,k-1|N}F_{k-1|N}^T- P_{k,k-1|N}^TF_{k-1|N}\}
\end{align}
\end{footnotesize}
where $  w1_{k|N}=X_{k|N}-f(X_{k-1|N})$.

\subsection{The Proposed DSDT Method for Estimating \textbf{Q}}
\label{DSDT}
\par We now estimate \textbf{Q} in a different way using the difference between the stochastic and dynamical trajectory (DSDT) method by following the extended EM method. The stochastic trajectory with the process noise can be approximated using the first order Taylor series expansion around a nominal point ($X_n$) as
\begin{align}
\label{e30} X_k&=f(X_{n_{k-1}})+f'(X_{n_{k-1}})(X_{k-1}-X_{n_{k-1}})+w_k
\end{align}
where $f'$ represents partial differentiation operation on f and thus $f'=\frac{\partial f}{\partial X}$. Consider the \\ dynamical trajectory ($Xd$) without the process noise defined as
\begin{align}
\nonumber Xd_{k}&=f(Xd_{k-1})\\
\label{e31} Xd_k&=f(X_{n_{k-1}})+f'(X_{n_{k-1}})(Xd_{k-1}-X_{n_{k-1}})
\end{align}
where and $Xd_0=X_0$. It is assumed that the nominal point ($X_n$) of both the above trajectories are close to the estimated dynamical trajectory ($X_{n_k}\approx Xd_{k|N}$) where $Xd_{k|N}=f(Xd_{k-1|N})$ and $Xd_{0|N}=X_{0|N}$. Subtracting Eq-\ref{e31} from Eq-\ref{e30} we get
\begin{align}
\nonumber X_k-Xd_{k}=&f'(Xd_{k-1|N})(X_{k-1}-Xd_{k-1|N}-Xd_{k-1}+Xd_{k-1|N})+w_k\\
\label{e20} w_k=&X_k-Xd_k-Fd_{k-1|N}(X_{k-1}-Xd_{k-1})
\end{align}
where the dynamical state Jacobian, $Fd_{k-1|N}=\frac{\partial f}{\partial X}_{|X=Xd_{k-1|N}}$.
\begin{align}
\nonumber w_kw_k^T&=X_kX_k^T-X_kXd_k^T-X_kX_{k-1}^TFd_{k-1}^T+X_kXd_{k-1}^TFd_{k-1}^T-Xd_kX_k^T+Xd_kXd_k^T\\
\nonumber &+Xd_kX_{k-1}^TFd_{k-1}^T-Xd_kXd_{k-1}^TFd_{k-1}^T-Fd_{k-1|N}X_{k-1}X_{k}^T+Fd_{k-1|N}X_{k-1}Xd_{k}^T\\
\nonumber &+Fd_{k-1|N}X_{k-1}X_{k-1}^TFd_{k-1}^T-Fd_{k-1|N}X_{k-1}Xd_{k}^TFd_{k-1}^T+Fd_{k-1|N}Xd_{k-1}X_{k}^T\\
\nonumber &-Fd_{k-1|N}Xd_{k-1}Xd_{k}^T-Fd_{k-1|N}Xd_{k-1}X_{k-1}^TF_{k-1|N}^T+Fd_{k-1|N}Xd_{k-1}Xd_{k}^TFd_{k-1}^T
\end{align}
Additionally apart from Eq-\ref{r1}, we have the following results
\begin{align}
\left.\begin{array}{r@{\;}l}
E[X_kXd_k^T|{Z_N}]&=E[X_k|Z_N]E[Xd_k^T|Z_N]=X_{k|N}Xd_{k|N}\\
E[Xd_kXd_k^T|{Z_N}]&=Xd_{k|N}Xd_{k|N}^T+Pd_{k|N}\\
E[Xd_kXd_{k-1}^T|{Z_N}]&=Xd_{k|N}Xd_{k-1|N}^T+Pd_{k,k-1|N}
\end{array}\right\}\label{r2}
\end{align}
where $Xd_{k|N}=f(Xd_{k-1|N})$ is the predicted dynamical state trajectory without the measurement and process noise using the estimated parameter, $\Theta_{N|N}$. Using Eq-\ref{r1} and Eq-\ref{r2} we get
\begin{align}
\nonumber E\left[w_kw_k^T|Z_N\right]=&  w2_{k|N}  w2_{k|N}^T+P_{k|N}+Fd_{k-1|N}P_{k-1|N}Fd_{k-1|N}-P_{k,k-1|N}Fd_{k-1|N}^T\\
\nonumber &-Fd_{k-1|N}P_{k,k-1|N}^T+Pd_{k|N}+Fd_{k-1|N}Pd_{k-1|N}Fd_{k-1|N}\\
\label{e33}&-Pd_{k,k-1|N}Fd_{k-1|N}^T-Fd_{k-1|N}Pd_{k,k-1|N}^T
\end{align}
where $w2_{k|N}=X_{k|N}-Xd_{k|N}-Fd_{k-1|N}(X_{k-1|N}-Xd_{k-1|N})$. Consider the following term,
\begin{align}
\nonumber &Xd_k-Xd_{k|N}=f(Xd_{k-1})-f(Xd_{k-1|N})\\
\nonumber &\approx f(Xd_{k-1|N})+Fd_{k-1|N}(Xd_{k-1}-Xd_{k-1|N})-f(Xd_{k-1|N})\\
\label{D1}&\approx Fd_{k-1|N}(Xd_{k-1}-Xd_{k-1|N})
\end{align}
Using Eq-\ref{D1}, we get the covariance of the dynamical trajectory as
\begin{align*}
Pd_{k|N}=&E[(Xd_k-Xd_{k|N})(Xd_k-Xd_{k|N})^T]\\
=&Fd_{k-1|N}Pd_{k-1|N}Fd_{k-1|N}^T
\end{align*}
where $Pd_{0|N}=P_{0|N}$ since $Xd_{0|N}=X_{0|N}$. The lag one covariance of dynamical trajectory is
\begin{align*}
Pd_{k,k-1|N}=&E[(Xd_k-Xd_{k|N})(Xd_{k-1}-Xd_{k-1|N})^T]=Fd_{k-1|N}Pd_{k-1|N}
\end{align*}
Substituting the value of $Pd_{k,k-1|N}$ and $Pd_{k|N}$ in Eq-\ref{e33} and using Eq-\ref{max2} we get
\begin{footnotesize}
\begin{align}
 \textbf{Q}=&\frac{1}{N}\sum_{k=1}^N \{  w2_{k|N}  w2_{k|N}^T+P_{k|N}+Fd_{k-1|N}P_{k-1|N}Fd_{k-1|N}^T-P_{k,k-1|N}Fd_{k-1|N}^T-P_{k,k-1|N}^TFd_{k-1|N}\}
\end{align}
\end{footnotesize}
%where $  w_{2_k}=X_{k|N}-X_{d_{k}}-F_{d_{k-1}}(X_{k-1|N}-X_{d_{k-1}})$.
If \textbf{Q} = 0 then X = $Xd$ and assuming that $P_{0|N}\approx$ 0, \textbf{R} can be estimated as
\begin{align}
\label{R4} {\textbf{R}}\approx \frac{1}{N}\sum_{k=1}^{N}{\left\lbrace (Z_k-h(Xd_{k|N}))(Z_k-h(Xd_{k|N}))^T\right\rbrace}
\end{align}

\subsection{Choice of \textbf{R}}
\label{sum1}
The choice of \textbf{R} for the next filter pass can utilize one appropriate among the many that are possible. Bavdekar et al. (2011) used the smoothed residue $Z_k-h(X_{k|N})$ for \textbf{R} estimation using extended EM method given by,
\begin{align}
\label{R1} \textbf{R}=&\frac{1}{N}\sum_{k=1}^{N}\left\lbrace(Z_k-h(X_{k|N}))(Z_k-h(X_{k|N}))^T+H_{k|N}P_{k|N} H_{k|N}^T\right\rbrace
\end{align}
 The choice of Mohamed and Schwarz (MS) for \textbf{R} estimation  based on the filtered residue is
\begin{align}
\label{R2} \textbf{R}&=\frac{1}{N}\sum_{k=1}^{N}{\left\lbrace (Z_k-h(X_{k|k}))(Z_k-h(X_{k|k}))^T+H_{k|k}{P}_{k|k}H_{k|k}^T\right\rbrace}
\end{align}
The choice of Myers and Tapley (MT) for  \textbf{R} estimation based on the innovation is
\begin{align}
\label{R3} {\textbf{R}}&=\frac{1}{N}\sum_{k=1}^{N}{\left\lbrace (Z_k-h(X_{k|k-1}))(Z_k-h(X_{k|k-1}))^T-H_k{P}_{k|k-1}H_k^T\right\rbrace}
\end{align}

Thus the above three equations use respectively smoothed, after update, and before updated states, the measurement and their corresponding covariances. All the measurement noise statistics innovations, filtered residue, smoothed residue are assumed to be zero mean. The smoothed residue is the best statistic for \textbf{R} estimation.
\subsection{Choice of \textbf{Q}}
\label{sum2}
The choice of \textbf{Q} for the next filter pass can utilize one appropriate among the many that are possible. Bavdekar et al. (2011) used the smoothed statistic $X_{k|N}-f(X_{k-1|N})$ for the \textbf{Q} estimation using extended EM method given by,
\begin{footnotesize}
\begin{align}
\label{Q1} \textbf{Q}=&\frac{1}{N}\sum_{k=1}^N \left\lbrace  w1_{k|N}   w1_{k|N}^T+P_{k|N}+F_{k-1|N}P_{k-1|N}F_{k-1|N}^T-P_{k,k-1|N}F_{k-1|N}^T-P_{k,k-1|N}^TF_{k-1|N}\right\rbrace
\end{align}
\end{footnotesize}
where $ w1_{k|N}=X_{k|N}-f(X_{k-1|N})$. 

The new DSDT statistic for \textbf{Q} (Section-\ref{DSDT}) is
\begin{footnotesize}
\begin{align}
\label{Q4} \textbf{Q}=&\frac{1}{N}\sum_{k=1}^N \left\lbrace  w2_{k|N}   w2_{k|N}^T+P_{k|N}+Fd_{k-1|N}P_{k-1|N}Fd_{k-1|N}^T-P_{k,k-1|N}Fd_{k-1|N}^T-P_{k,k-1|N}^TFd_{k-1|N}\right\rbrace
\end{align}
\end{footnotesize}
where $w2_{k|N}=X_{k|N}-Xd_{k|N}-Fd_{k-1|N}(X_{k-1|N}-Xd_{k-1|N})$. Mohamed and Schwarz (MS) used innovations and  gain for estimating \textbf{Q} given by,
\begin{align}
\label{Q2} \textbf{Q}&=K_N\left\lbrace \frac{1}{N}\sum_{k=1}^N (Z_k-h(X_{k|k-1}))(Z_k-h(X_{k|k-1}))^T\right\rbrace K_N^T
\end{align}
The choice of Myers and Tapley (MT) for \textbf{Q} is $w3_{k|k}=X_{k|k}-X_{k|k-1}$ and is given by,
\begin{align}
\label{Q3} {\textbf{Q}}&=\frac{1}{N}\sum_{k=1}^{N}{\left\lbrace   w3_{k|k}  w3_{k|k}^T-\left(F_{k-1}{P}_{k|k-1}F_{k-1}^T-P_{k|k}\right)\right\rbrace}
\end{align}
All the process noise samples, $w1_{k|N},w2_{k|N}$ and $w3_{k|k}$ are assumed to be zero mean. We note that the smoothed statistics $w1_{k|N}$ and $w2_{k|N}$ provide very close results and are the best for the \textbf{Q} estimation.

Much of the earlier work used simulated data to tune the filter off line for obtaining the statistics to be used later for on line and real time applications. While this is correct one important feature has been forgotten in such approaches. The convergence of any technique even in simulation studies is no guarantee for a proper solution to the problem. Even the simple case of a linear fit to a set of data many variants tend to different results (TR/EE2015/401\cite{report}). Hence we still lean on simulation studies with exact solutions being available to the analyst. Hence presently the filter methods have been applied firstly to a very simple spring, mass, and damper system for which when only the measurement noise exists and the process noise does not exist (\textbf{Q} = 0). For such a situation exact reference results are derivable by using the Newton Raphson (NR) technique (Ananthasayanam et al. 2001) and these can be compared with filter generated results. Subsequently when the process noise in included in the system it is necessary to look for the consistency based on a comparison between the injected measurement and process noise sequence and their statistics. Further many other cost functions based on balancing the states and measurement equations that are introduced help to move from deceptive to decisive solutions. The estimation of \textbf{R} and \textbf{Q} is possible provided one utilizes appropriate choice of the estimation `statistics' based on many quantities that arise in the filter operation like the pre and post filter states as well as the ones derivable from the measurements such as the innovation, filtered residue, smoothed residue and their covariances.

The above helped to guide us in the choice of appropriate initial filter statistics namely $\mathbf{X_0, P_0,}\Theta$, \textbf{Q} and \textbf{R}. In particular after the first filter pass (using some initial guess  statistics) through the data which is a must it has guided the way the filter statistics have to be updated from the second iteration onwards. It helps to answer the questions like what should be the $\mathbf{P_0}$ for the states and the parameters be derived from the value at the end of the pass to be used in the next iteration, should they be full, diagonal, or even zero, whether the \textbf{Q} has to be injected into the state and/or the augmented parameter states. Since the reference NR estimates of the parameters, the Cramer Rao bound (CRB), and the measurement noise are available it has been possible to settle such questions and in particular the CRB played a crucial role to guide the choices for the above. Generally the Kalman filter provides fairly acceptable estimates for the parameters but unless the tuning is good it is very difficult to match the CRBs from the EKF with the NR values.

\subsection{Adaptive Tuning and the present Reference Recursive Recipe}

The different methods and options for filter tuning forms a part of sensitivity study which are,

\begin{enumerate}
\item $\mathbf{P_0}$ can be estimated by Scale up, IIM  or by Smoothing ($P_{0|N}$).
\item Options for $\mathbf{P_0}$ which can be split as cov([State-S;Parameter-P]) are

\begin{enumerate}
\item  Reference Matrix-[0,0;0,\checkmark],
\item  Diagonal matrix - [\checkmark,0;0,\checkmark],
\item  Full matrix - [\checkmark,\checkmark;\checkmark,\checkmark].
\end{enumerate}

The checkmark (\checkmark) represent a nonzero value at the indicated position. The `cov (.)' represents covariance matrix.
\item The process noise \textbf{Q} can be estimated by using Eq-\ref{Q1}, Eq-\ref{Q4}, Eq-\ref{Q2} or Eq-\ref{Q3}.
\item Options for \textbf{Q} = cov([S;P]) also include the same options used for $\mathbf{P_0}$ except that the refrence matrix is [\checkmark,0;0,0].
\item The measurement noise \textbf{R} can be estimated by using Eq-\ref{R1}, Eq-\ref{R2} or Eq-\ref{R3}.
\end{enumerate}

The following steps explain the recursive or iterative algorithm for tuning the EKF.

\begin{enumerate}

\item Given the system model and the measurements the first iteration of EKF is carried out with guess values of $\mathbf{X_0}$, $\mathbf{P_0}$, $\Theta$, \textbf{R} and \textbf{Q}.

\item Run the extended RTS smoother using the filtered data to get the smoothed state estimate $X_{k|N}$ and the corresponding smoothed covariance $P_{k|N}$.

\item The $\mathbf{P_0}$ can be estimated by Scale up (Eq - \ref{P2}), IIM (Eq - \ref{P1}) or the smoothed ($P_{0|N}$) which will have to be scaled up and modified for obtaining proper results.

\item The measurement and process noise covariance can be estimated by any of the options as discussed in Section-\ref{sum1} and \ref{sum2}.

\item EKF is run using the updated estimates of $\mathbf{X_0}$, $\mathbf{P_0}$, $\Theta$, \textbf{Q} and \textbf{R} at the beginning of further iterations until statistical equilibrium is reached.

\item Different cost functions (\textbf{J1} to \textbf{J8}) are checked for convergence.

\item Many simulation runs (say 50) are carried out by varying the injected measurement ($v$) and process noise ($w$) sequences.

\end{enumerate}

For the \textbf{Q} = 0 case the value of \textbf{Q} is set at $10^{-10}$ or lower for all iterations to help the filter that would otherwise generate a pseudo \textbf{Q} and then slowly grind it to zero in hundreds of iterations. For the \textbf{Q} $>$ 0 case if any of the states is known to have \textbf{Q} = 0 then it can be set at $10^{-10}$ or lower. For \textbf{Q} = 0 case one can even estimate \textbf{R} by ignoring the second order terms (assuming \textbf{P}$\rightarrow$ 0). It is of interest to note that for \textbf{Q} $>$ 0 case unless the second order terms of the filter output covariance terms are also included in (i) the estimate for \textbf{R }and \textbf{Q} using the EM option and (ii) the estimate for \textbf{R} using the EM together with \textbf{Q} using the DSDT option the estimates for \textbf{R} and \textbf{Q} do not converge to the proper value. The different adaptive approaches analysed in the paper are provided in Table-\ref{conc}. Based on a comparative study the proposed RRR is as below.

\begin{table}[h]
\begin{center}
\caption*{\textbf{The proposed Reference Recursive Recipe (RRR)}}{}
\begin{tabular}{| c | c |  }
\hline
\textbf{Q} = 0 &  \textbf{Q} $>$ 0 \\ \hline
\makecell{ $\mathbf{X_0}$ : Given or $X_{0|N}$  \\ $\Theta$ : $\Theta_{N|N}$  \\ $\mathbf{P_0}$ : Scaled up-[0,0;0,\checkmark]\\\textbf{Q} : $10^{-10}$-[\checkmark,0;0,0] \\ \textbf{R} : EM-diag} &
\makecell{$\mathbf{X_0}$ : Given or $X_{0|N}$ \\ $\Theta$ : $\Theta_{N|N}$  \\ $\mathbf{P_0}$ : Scaled up-[0,0;0,\checkmark]\\\textbf{Q} : EM/DSDT-[\checkmark,0;0,0] \\ \textbf{R} : EM-diag} \\ \hline
\end{tabular}
\end{center}
\end{table}

We call this as a reference and not as a standard since improvements could be made later when such a recursive filtering approaches match the solutions provided by optimization techniques like NR or other involved ones including \textbf{Q}. The $\mathbf{X_0}$ in all cases is either given or obtained by the smoother. The smoothed initial estimate which can be split as $X_{0|N}$ =  $(x_{0|N},\Theta_{0|N})^T$ including both state and parameter. The estimated parameter is taken as $\Theta_{N|N}$ obtained from $X_{N|N}=[x_{N|N},\Theta_{N|N}]$ with covariance $P_\Theta$ obtained at the end of the final filter pass over the data, $P_{N|N}$ = $[ P_{xx}, P_{x\Theta}; P_{\Theta x}, P_\Theta]$.

Generally mathematical treatments implicitly assume unlimited data. However, in practice we have to deal with finite lengths of data and so need stable convergence properties reasonably independent of starting estimates. In any field of science or technology one needs at times stability and at other times sensitivity. In the present RRR there is stability in the procedure commencing from different reasonable initial values converging to the same result for a given measurement data. Should the measurement data change then the converged parameters also change (!) thus showing sensitivity.

\section{Simulation Study of a Spring, Mass and Damper System}
\label{simulation}
Consider the SMD system with weak nonlinear spring constant in the continuous time (t) state space form given by
\begin{align*}
%\left.\begin{array}{r@{\;}l}
\dot{x}_1(t)&=x_2(t)\\
\dot{x}_2(t)&=-\Theta_1x_1(t)-\Theta_2x_2(t)-\Theta_3x_1^3(t)
%\end{array}\right\}\label{e24}
\end{align*}
where $x_1$, $x_2$ are the displacement and velocity state with initial conditions 1 and 0 respectively. The $\dot{x}$ represents differentiation with respect to time (t). The unknown parameter vector is $\Theta=[\Theta_1,\Theta_2,\Theta_3]^T$ with the true values $\Theta_{true}=(4,0.4,0.6)^T$. The $\Theta_3$ is a weak parameter since its value do not affect the system dynamics much. The complete state vector, X=$[x_1,x_2,\Theta_1,\Theta_2,\Theta_3]^T$ of size $(n+p)\times 1$. The measurement equation is given by
\begin{align*}
Z_k=HX_k+v_k
\end{align*}
where H=$\begin{bmatrix}
1 & 0 & 0 & 0 & 0 \\ 0 & 1 & 0 & 0 & 0
\end{bmatrix}$ is the measurement matrix of size $m\times (n+p)$ where $m=n=2$ and $p=3$. The values of the noise variances are R = diag(0.001,0.004) and Q = diag(0.001,0.002) where `diag' is the diagonal operator as used in MATLAB\textsuperscript{\textregistered}. All the figures are presented for only one simulation run to prevent cluttering. In the SMD system study, the guess value of $\mathbf{P_0}$ chosen is $10^{-1}$ for all states which is assumed to be a diagonal matrix in the first iteration. The guess value of \textbf{Q} chosen is $10^{-1}$ for all states and zero for the augmented parameters. The guess value of \textbf{R} chosen is $2^{-1}$ for all measurement channels. The initial parameters are chosen to be within $\pm 20 \%$ error. A total of N = 100 measurement data are simulated with the time varying from 0 to 10 seconds in very small steps of $\delta$t = 0.1 s. For zero process noise case, the maximum number of iteration is set to 20 over $n_s$=50 simulations and for non zero process noise case it is set to 100 over 50 simulations for obtaining generally four digit accuracy (though not necessary) in the results as presented in the Table-\ref{tbsmd} and \ref{tbsmdQ}. In the present reference procedure it was noticed that generally even if the initial state covariance, initial process and measurement noise variances are varied over a wide range of powers from -3 to +3 together with the initial parameter values being set to zero one can reach the same estimation results for a given data. Thus there is stability with respect to far off initial conditions but sensitivity in the estimates with different data. Such studies show that the present reference recursive recipe leads to a non diverging, and consistent filter performance over many simulations and provides better results when compared to earlier approaches. The convergence of the following quantities through the iterations are analyzed.

\begin{enumerate}
\item The parameter estimates $\Theta$ and their covariances $P_{\Theta}$.
\item The noise covariance of \textbf{Q} and \textbf{R}.
\item The state dynamics without measurement and process noises based on the estimated parameter after the filter pass through the data Xd, the prior state X-, the posterior state X+, the smoothed state Xs and the measurement Z.
\item The sample innovation, filtered residue and the smoothed residue along with their bounds which is the square root of the predicted covariances given respectively by ($\textbf{R}+H_kP_{k|k-1}H_k^T$), ($\textbf{R}-H_{k|k}P_{k|k}H_{k|k}^T$) and ($\textbf{R}-H_{k|N}P_{k|N}H_{k|N}^T$) by the filter.
\item The estimated measurement and process noise samples as well as their autocorrelations.
\item The cost functions (\textbf{J1-J8}) after the final convergence.
\end{enumerate}

\subsection{Discussion of the Results}

The system under consideration is first solved using the recursive reference recipe (RRR). Later other sensitivity studies were carried out using other possible variations for $\mathbf{P_0}$, \textbf{Q} and \textbf{R}. Such studies lead to the conclusion that the reference recipe is just about the best possible and others may not always help the filter operation without divergence and even if they did may not lead to results as good as the reference recipe. The tabulated results are averaged over 50 simulations which includes the following
\newpage
\begin{itemize}
\item $\Theta$ ratio EKF/NR is the ratio of EKF estimated parameter ($\Theta_{N|N}$) to that of the parameter estimated by NR method.
\item $\Theta$ ratio EKF/True is the ratio of EKF estimated parameter ($\Theta_{N|N}$) to that of $\Theta_{true}$.
\item CRB ratio is the ratio of the square root of the parameter covariance ($P_{\Theta}$) estimated by EKF to that of the CRB estimated by the NR method.
\item $
\text{Consistency ratio = } {\sigma_{\Theta}}/{SIGMA}_{avg}$ \\
\begin{flushleft}
$
 \text{ where } \sigma_{  \Theta}=\left[\frac{1}{n_s}\sum\limits_{s=1}^{n_s}{(  \Theta^s-\bar \Theta)^2}\right]^{\frac{1}{2}} \text{, }
SIGMA_{avg}=\frac{1}{n_s}\sum\limits_{s=1}^{n_s}{\sqrt{P^s_{\Theta}}}$, $n_s$ is the total number of simulations, s is the simulation number, $\bar \Theta$ is the sample mean of the estimated parameters.
\end{flushleft}
\item Spread factor is a measure of percentage spread seen in the parameter estimates using both first and second order moments which is defined as
\begin{equation*}
\text{Spread factor = }\left[\frac{1}{n_s}\sum_{s=1}^{n_s}{\sqrt{(\Theta-  \Theta^s)^2 + P_{  \Theta}^s}}\right]\times \frac{100}{|\Theta|}
\end{equation*}
\item \textbf{R} ratio EKF/True is the ratio of the EKF estimated \textbf{R} to that of the true value of \textbf{R}.
\item \textbf{R} ratio EKF/NR is the ratio of the EKF estimated \textbf{R} to that estimated by NR method.
\item \textbf{Q} ratio is the ratio of the EKF estimated \textbf{Q} to that of the true \textbf{Q}.
\item The mean ($\mu$) and standard deviation ($\sigma$) of the cost functions (\textbf{J1-J8}) are over many simulations ($n_s$=50).
\end{itemize}

%The Table-\ref{tbsmd} shows the results for the Q = 0 case. The appropriate reference results are shown inside the thick block. The value of Q is set to $10^{-10}$ since otherwise the filter takes more than hundreds of iterations for convergence. One can note that the parameter estimates, the ratio of \textbf{R}, the cost functions are fairly comparable in the different cases. However it is the ratio of the filter estimated CRBs, (whose effect is amplified and shown) by the consistency ratio and the spread factor that differ from the corresponding NR estimates. This feature indicates that the reference procedure with its choice of $\mathbf{P_0}$ is about the best among all possible options. When the smoothed $\mathbf{P_0}$ is used without scaling up the CRB ratio decreases due to the continuously decreasing $\mathbf{P_0}$ with iterations leads to incorrect consistency and spread factors in spite of good \textbf{R} ratio and cost function values. The formulae for \textbf{R} can use smoothed residue, filtered residue, and innovations with and without second order terms. In addition we can add as another estimate for \textbf{R} the difference between the measured and predicted dynamics. It turns out that when \textbf{Q} = 0 all the above seven options lead to statistically the same results.

\subsubsection{Without Process noise (\textbf{Q} = 0)}

\begin{figure}[t]
\includegraphics[width=6in,height=2.5in]{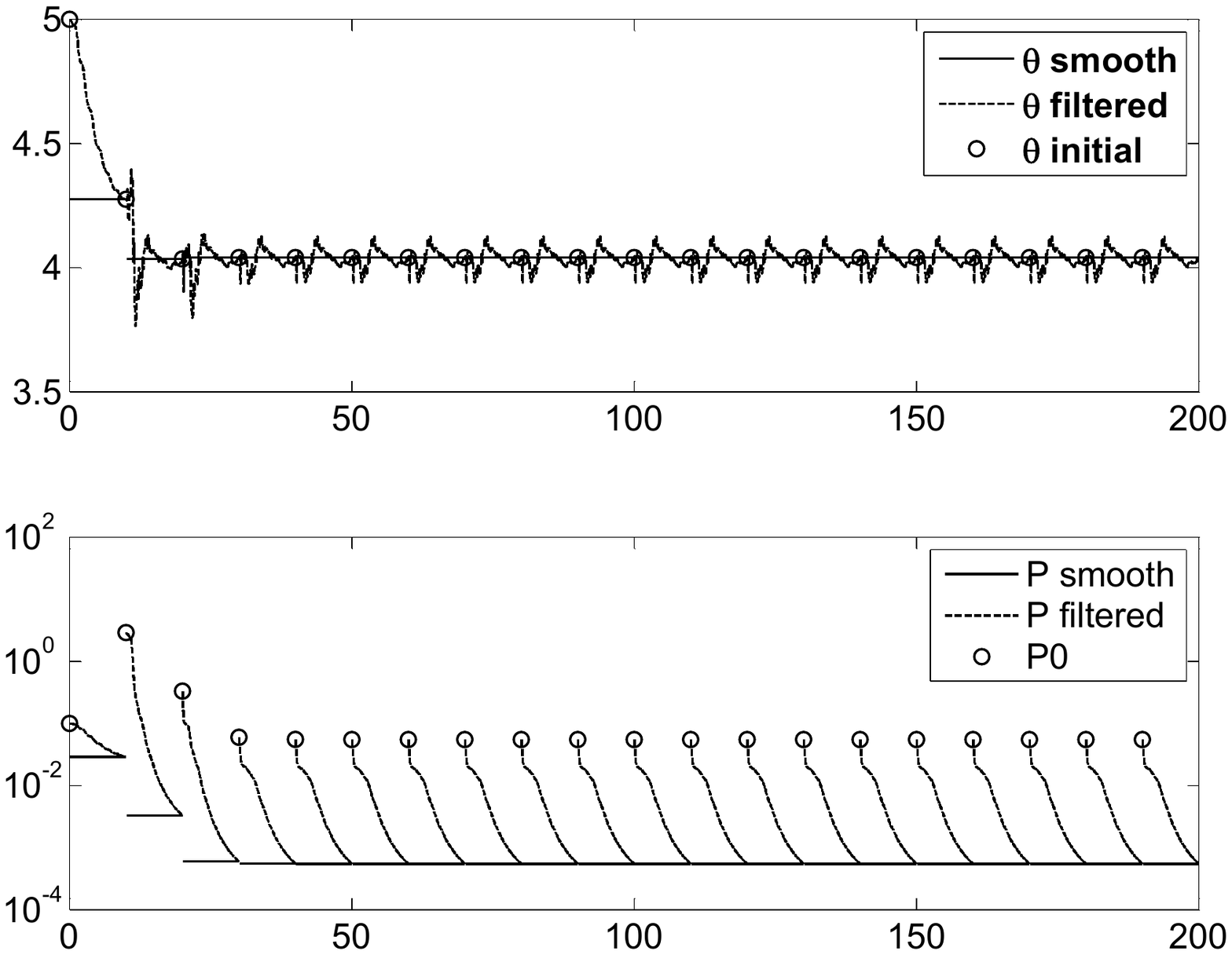}
\caption{Variation of $\Theta_1$ and $P_{\Theta_1}$ with cumulative time and iterations (\textbf{Q}=0)}
\label{theta1}
\end{figure}

\begin{figure}[h]
\includegraphics[width=6in,height=2.5in]{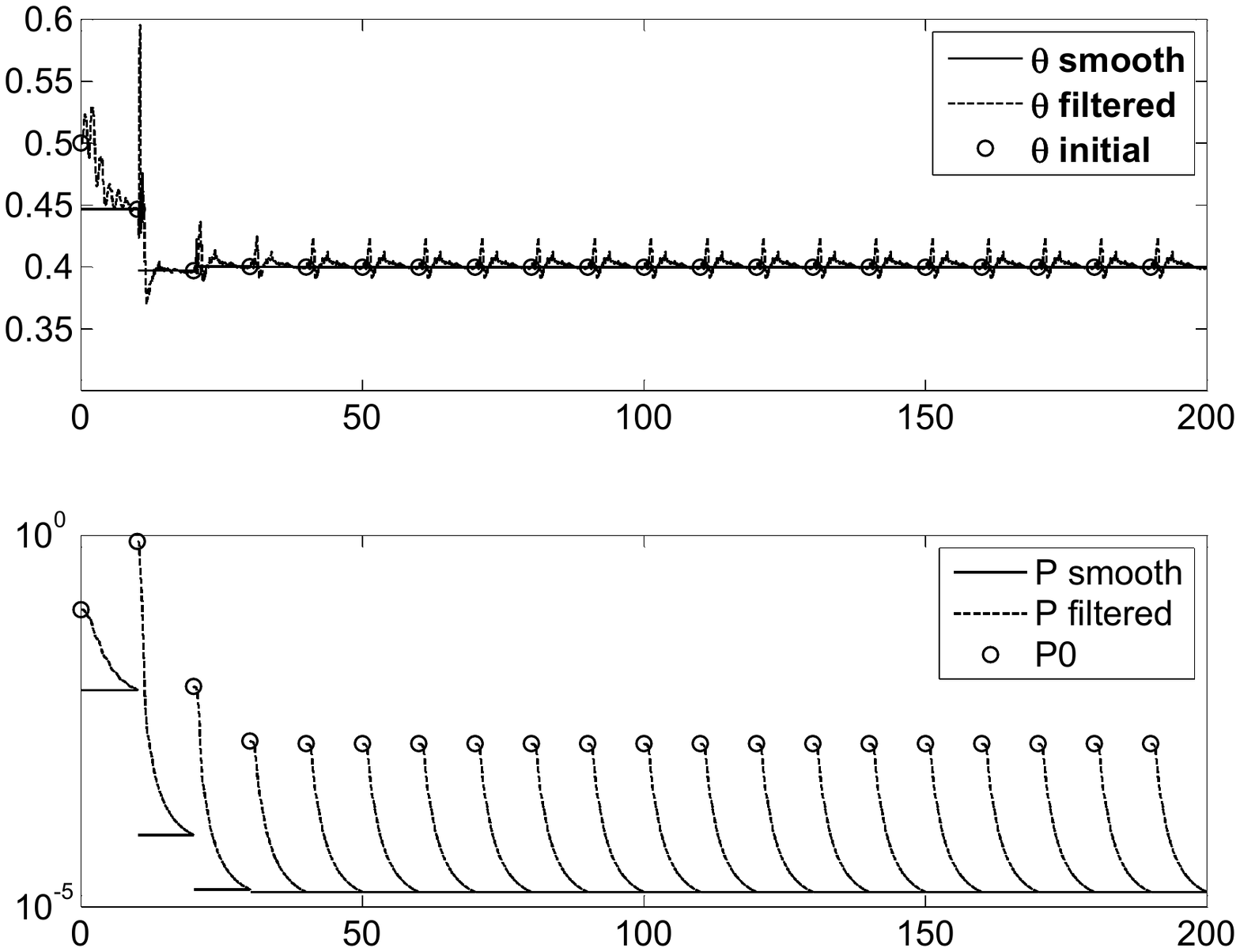}
\caption{Variation of $\Theta_2$ and $P_{\Theta_2}$ with cumulative time and iterations (\textbf{Q}=0)}
\label{theta2}
\end{figure}

\begin{figure}[h!]
\includegraphics[width=6in,height=2.5in]{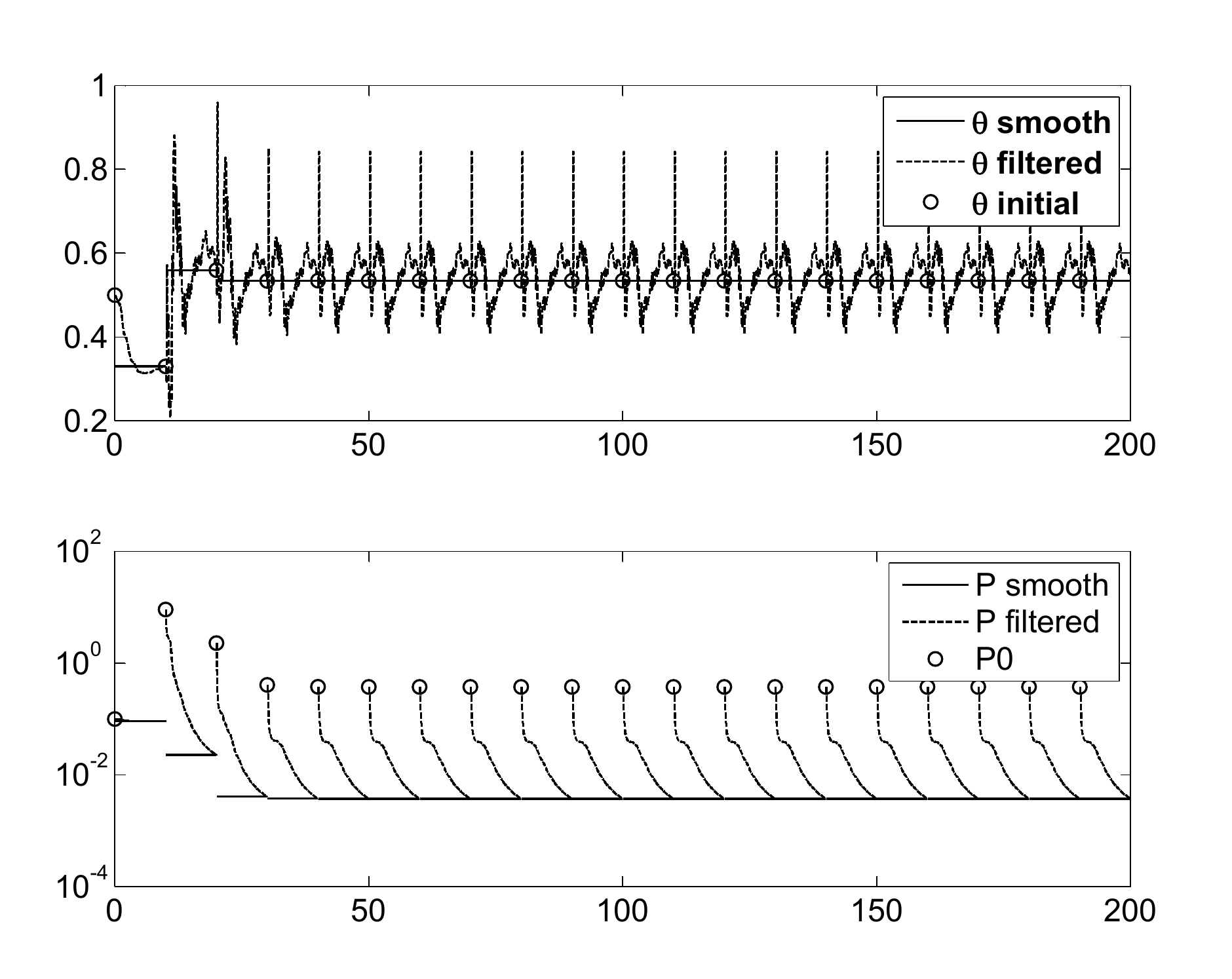}
\caption{Variation of $\Theta_3$ and $P_{\Theta_3}$ with cumulative time and iterations (\textbf{Q}=0)}
\label{theta3}
\end{figure}

The Table-\ref{tbsmd} and \ref{tbsmdb} show the results for the \textbf{Q} = 0 case. The appropriate reference results are shown inside the thick block. The later rows present results with different options. One can note that the parameter estimates, the ratio of \textbf{R}, and the cost functions are fairly comparable for different options. However it is the ratio of the filter estimated CRBs, (whose effect is amplified and shown) by the consistency ratio and the spread factor that differ from the corresponding NR estimates. This feature indicates that the reference procedure with its choice of $\mathbf{P_0}$, $\Theta$, \textbf{Q} and \textbf{R} updates is about the simplest and best among all possible options. The Table-\ref{tbsmdb} shows the effect of using simply the smoothed $\mathbf{P_0}$ without any scaling but using possible options. Even in this process noise free case due to the absence of scaling up the $\mathbf{P_0}$, the CRB ratio decreases due to the continuously decreasing smoothed $\mathbf{P_{0|N}}$ with iterations and leads to incorrect consistency and spread factors in spite of deceptively good \textbf{R} ratio and cost function values. The above results show the importance of proper scaling and trimming the $\mathbf{P_0}$. When \textbf{Q} = 0 the estimate for \textbf{R} can use smoothed residue, filtered residue, and innovations with and without second order terms. In addition another estimate for \textbf{R} can be the difference between the measured and predicted dynamics. It turns out that for process noise free case all the above seven options lead to statistically the same results. The Figs. \ref{theta1}, \ref{theta2}, and \ref{theta3} show the variation of the estimated initial parameters and their variances through 20 iterations using the RRR (\textbf{Q} = 0). The x-axis for the above three plots is the cumulatively increasing time with iterations. For example in the simulated SMD system, the time instants vary from 0 to 10 seconds in the first iteration in small time steps of 0.1 seconds. The second iteration have time instants varying from 10 to 20 seconds, the third iteration with 20 to 30 seconds and so on till 200 seconds for the $20^{th}$ iteration. The parameter and the uncertainty reach almost their final estimated values with three digit accuracy in about two and five iterations respectively.

The correlation coefficient matrix is $C_{ij}=\frac{d_{ij}}{\sqrt{d_{ii}\times d_{jj}}}$ where $d_{ij}$ is the $i^{th}$ row $j^{th}$ column element of the parameter covariance matrix $P_{\Theta}$ of size p$\times$ p estimated by the EKF at the last time instant of the final iteration. For the spring, mass, and damper system a typical value for one data set with zero process noise is
\begin{center}
C =
$\begin{bmatrix}
 1.0000  & -0.0024  & -0.9387\\
-0.0024  &  1.0000  &  0.1193\\
-0.9387  &  0.1193  &  1.0000
\end{bmatrix}$
\end{center}

The forces that balance the inertial force are dependent on velocity and displacement. Over a range of displacement that is excited the nonlinear spring constant ($\Theta_3$) can be estimated only with high correlation with the linear spring constant ($\Theta_1$). If the range of displacement is increased then the above correlation reduces. However the damping coefficient ($\Theta_2$) is estimated with modest correlation with other parameters since it is driven by velocity than the displacement.

\subsubsection{With Process noise (\textbf{Q} $>$ 0)}

\begin{figure}[t]
\includegraphics[width=6in,height=2.5in]{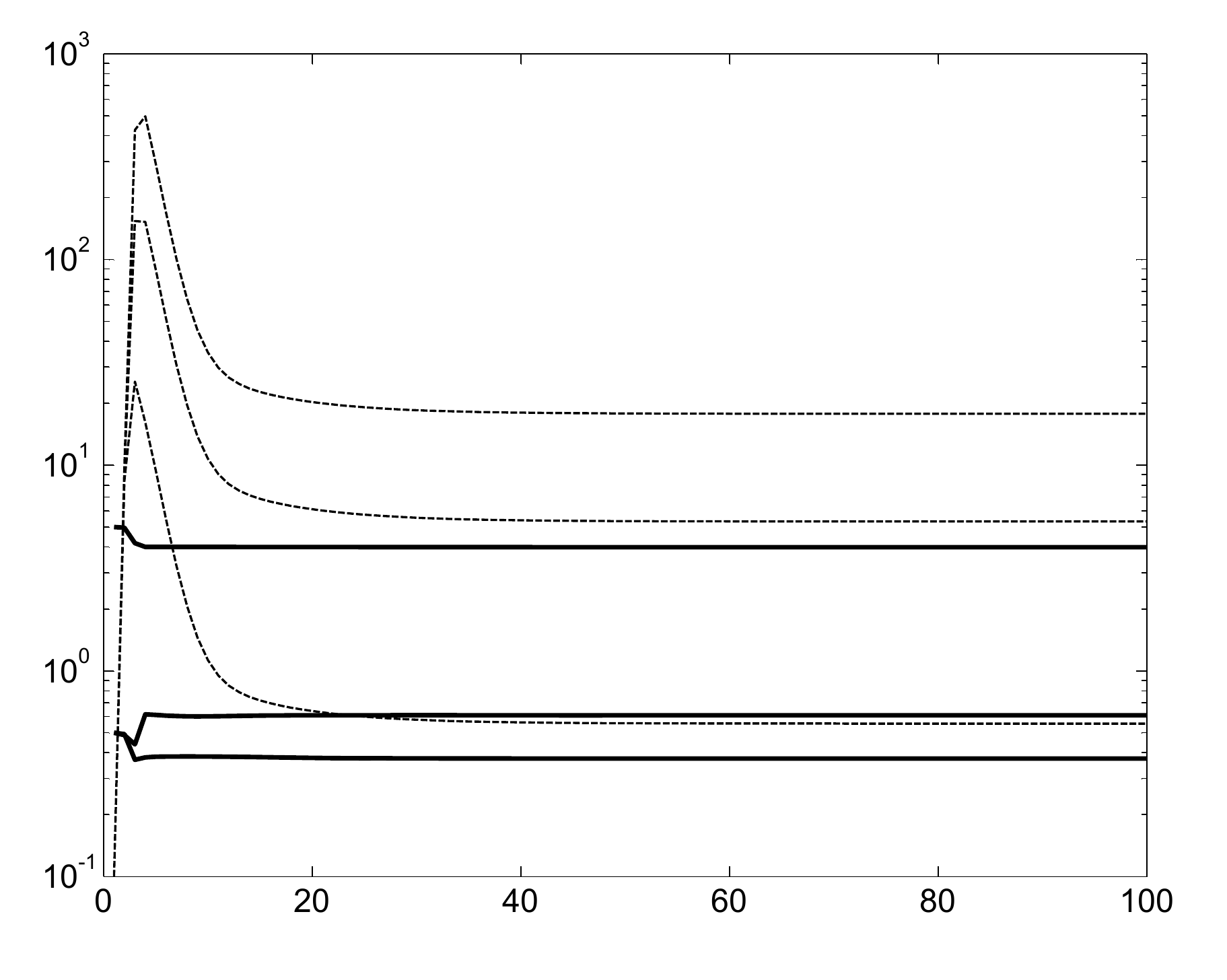}
\caption{Variation of initial parameters $\Theta_0$(continuous) and its $\mathbf{P_0}$(dashed) with iterations}
\label{smdQ_P0}
\end{figure}

\begin{figure}[h]
\includegraphics[width=6in,height=2.5in]{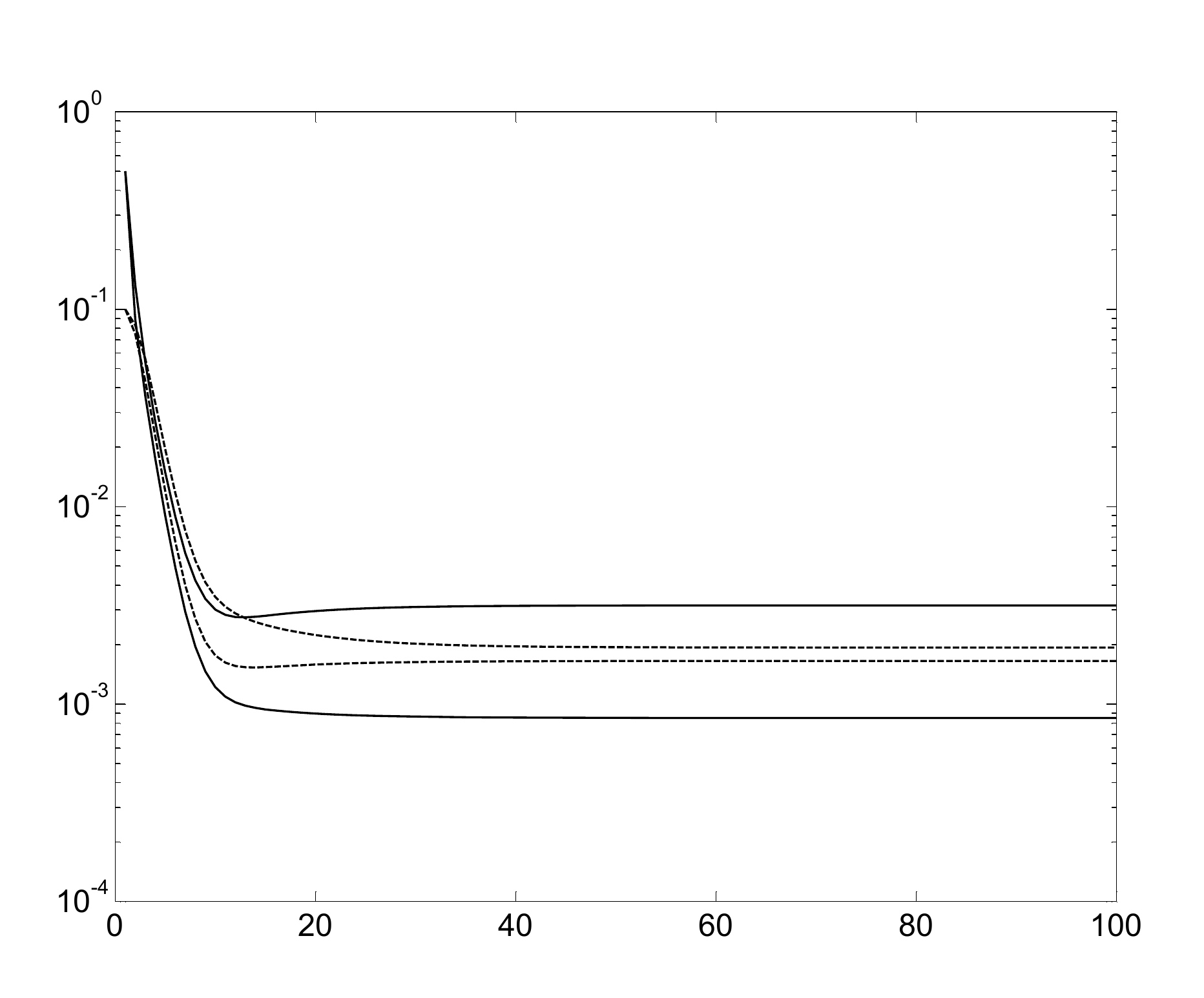}
\caption{Variation of \textbf{Q} (dashed) and \textbf{R} (continuous) with iterations}
\label{smdQ_R}
\end{figure}

\begin{figure}[b]
\includegraphics[width=6in,height=2.5in]{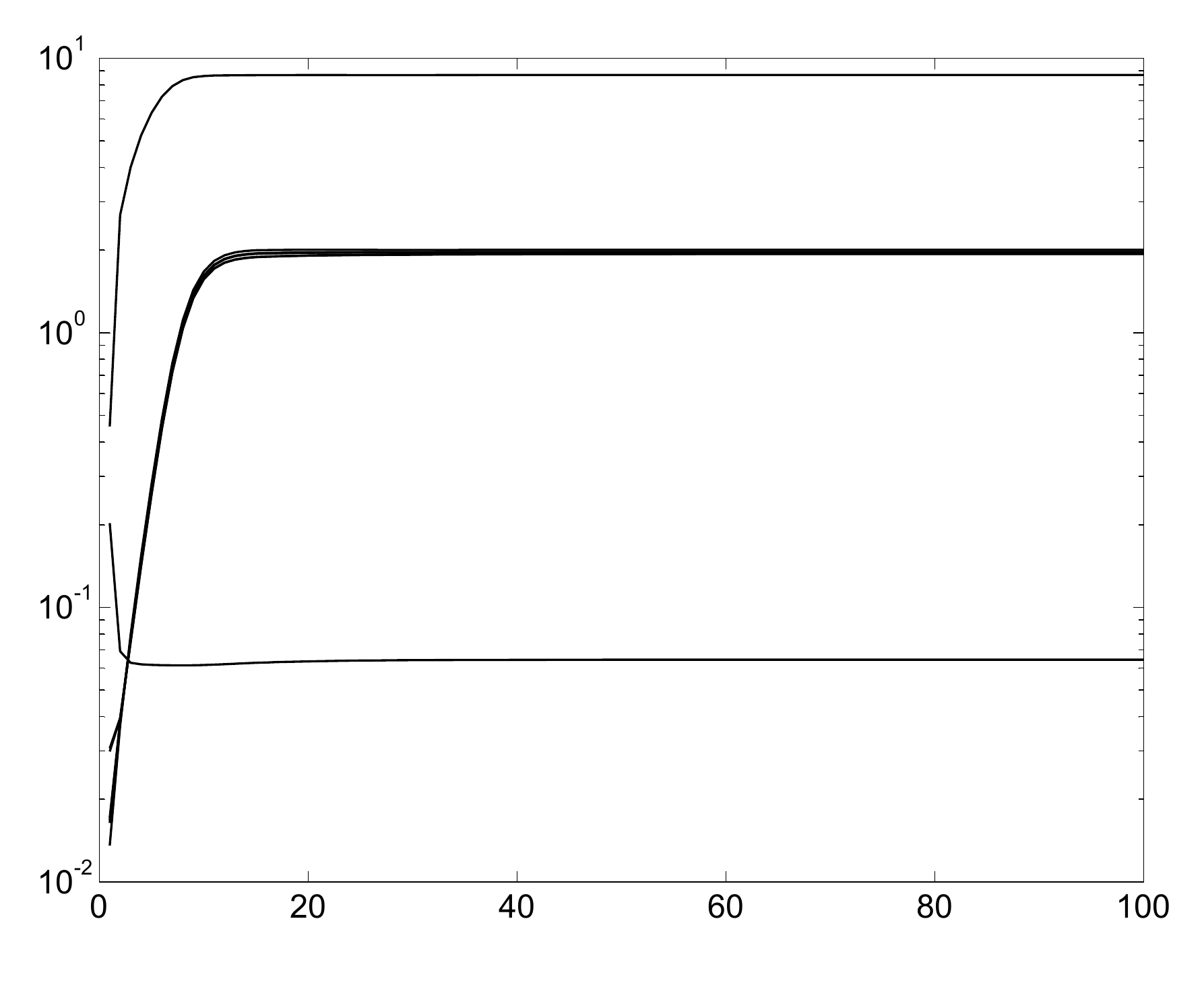}
\caption{Variation of different costs (\textbf{J1-J8}) with iterations}
\label{smdQ_J}
\end{figure}

The Figs.\ref{smdQ_P0}, \ref{smdQ_R}, and \ref{smdQ_J} show respectively the variation of (i) the initially estimated $\Theta$ and its variance, (ii) \textbf{R} and \textbf{Q} and (iii) the different cost functions (\textbf{J1-J8}) through 100 iterations for the RRR (\textbf{Q} $>$ 0) case. The cost functions \textbf{J1-J3} correspond to the number of measurement (m=2). The \textbf{J4 }in absence of process noise would correspond to the trace of the measurement noise \textbf{R}. The \textbf{J5} is the negative log likelihood cost whose absolute value is shown in the plot. In presence of process noise, \textbf{J6-J8} correspond to the number of states (n=2).

It has been further extensively checked in the report, TR/EE2015/401 (2015) that the variation of the sample innovations, filtered residue and smoothed residue are consistent with the square root of their filter predicted variances ($\pm \sigma$ bound). In the EKF approach since most of the quantities are Gaussian or approximated as quasi Gaussian and one would expect all the above quantities are close to being Gaussian and hence around one third of the total sample points to be outside the  $\pm \sigma$ bound. Similarly the injected and estimated measurement noise distributions during the final iteration were very close to each other as also their autocorrelations ideally expected to be close to the  Kronecker delta function which provides confidence in the proposed filter algorithm.

\begin{figure}[h]
\includegraphics[width=6in,height=4.0in]{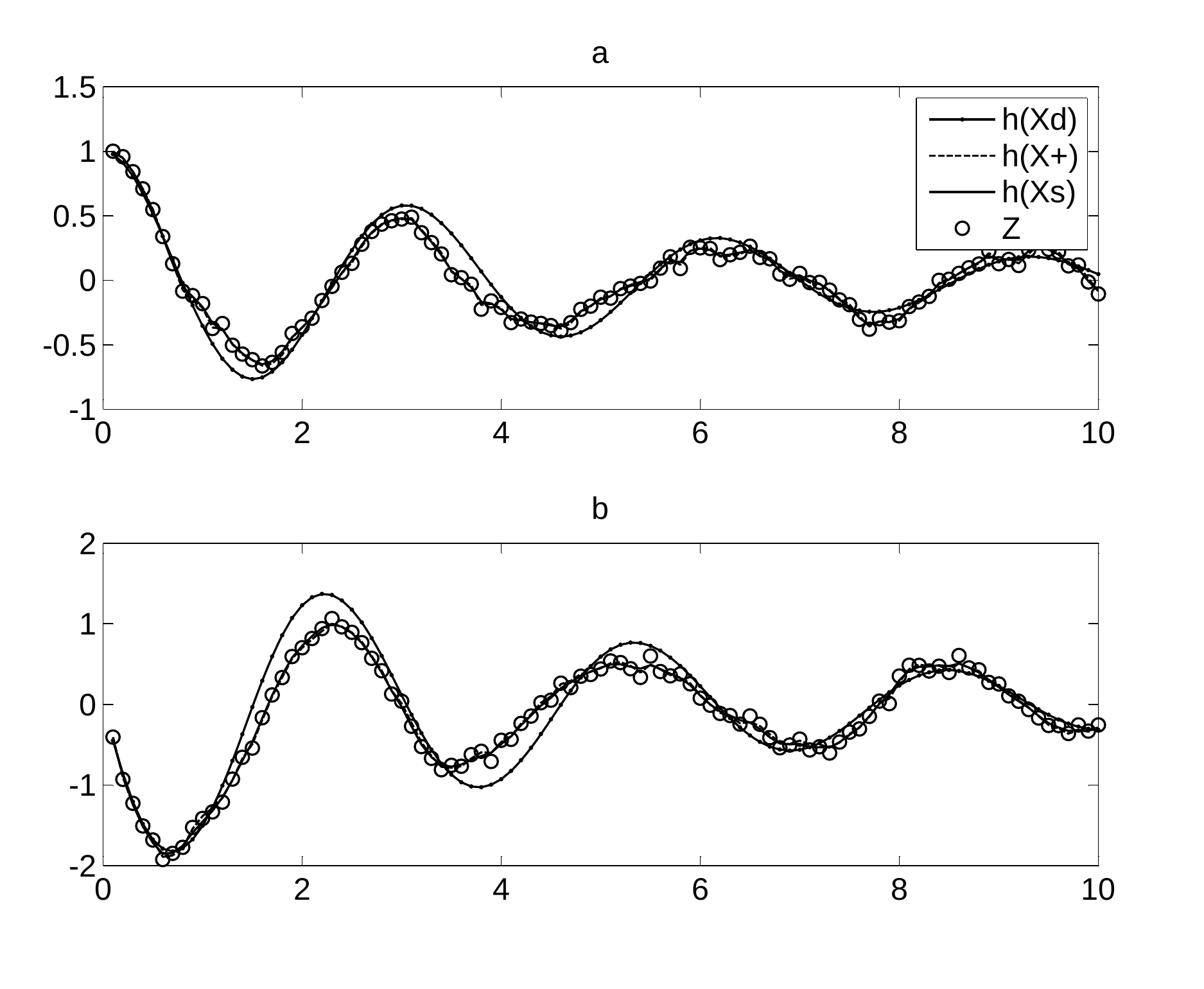}
\caption{Comparison of the predicted dynamics h(Xd), posterior h(X+), smoothed h(Xs)\\  and the measurement Z corresponding to the (a) displacement  (b) velocity  }
\label{smdQ_h}
\end{figure}

The Fig.\ref{smdQ_h} show the predicted dynamics, filtered and smoothed estimate at the last iteration. There is an expected mismatch in the estimated dynamics (without the effect of \textbf{R} and \textbf{Q}) and the measurements made on the wandering dynamics indicating the presence of process noise. The next Figs. \ref{MS_SMD_Q} to \ref{EM_SMD_J} show the variation of the noise \textbf{R} and \textbf{Q} estimates and the cost functions \textbf{J1} to \textbf{J8} for MS, MT, and Bavedkar et al approaches. The important features to be noted are the MS though it converges leads to inaccurate estimates for the noise statistics and cost functions. The MT and Bavedkar et al approaches have in general no systematic variation and convergence of the estimates. The Table-\ref{tbsmdQ} shows the reference case in the second row the cost functions \textbf{J1} to \textbf{J3} and \textbf{J6} to \textbf{J8} are close to their expected values. The third row shows by using the smoothed $\mathbf{P_0}$ without scaling and no stopping condition, the low spread factor is deceptive since there is no consistency in the parameter estimates with its covariance.

\begin{center}
\begin{table}[h]
\caption{Different Adaptive Methods for Comparative Analysis}
\label{conc}
\resizebox{\textwidth}{!}{\begin{tabu}{|c|c|c|c|c|}
\hline
Method	& Options for $\mathbf{P_0}$	& Options for \textbf{Q}	& Options for \textbf{R}	& Remarks \\ \hline

\makecell{NR\\(See TR/EE2015/401)}	& Not applicable &	Not applicable	& \makecell{Using $\left(Z - h(X_{\Theta})\right)$ }   &	\makecell{Applicable only \\for \textbf{Q}=0 case}. \\ \hline

\makecell{Reference\\ (Present)} & 	\makecell{Scaled up $P_{N|N}$ (Eq.\ref{P1}) \\ trimmed to [0,0;0,\checkmark] } & \makecell{EM Algorithm (Eq.\ref{Q1})\\ or DSDT (Eq.\ref{Q4})\\ trimmed to [\checkmark, 0; 0, 0]} & \makecell{Smoothed Residue \\$\left(Z_k - h(X_{k|N})\right)$ Eq.\ref{R1}} & \makecell{Statistical \\equillibrium and \\CRBs  are achieved}  \\ \hline

\makecell{IIM (Similar  to \\Gemson's method)} &\makecell{IIM (Eq.\ref{P2}) \\ trimmed to [0,0;0,\checkmark] } & \makecell{MT Method (Eq.\ref{Q3})\\ trimmed to [\checkmark, 0; 0, 0]} & \makecell{Innovations \\$\left(Z_k - h(X_{k|k-1})\right)$ Eq.\ref{R3}} & \makecell{Initial \textbf{R} should be close\\ to the true value.}  \\ \hline

\makecell{Bavdekar et al.\\ method} & 	\makecell{Smoothed $P_{0|N}$ (Eq.\ref{P3})} & \makecell{EM Algorithm (Eq.\ref{Q1})\\ with full matrix} & \makecell{Smoothed Residue \\$\left(Z_k - h(X_{k|N})\right)$ Eq.\ref{R1}} & \makecell{Cost functions diverge\\ after some iterations.\\ CRBs are not achieved.}  \\ \hline

\makecell{MT Method*} & \makecell{Scaled up $P_{N|N}$ (Eq.\ref{P1}) \\ trimmed to [0,0;0,\checkmark] } & \makecell{MT Method (Eq.\ref{Q3})\\ trimmed to [\checkmark, 0; 0, 0]} & \makecell{Innovations \\$\left(Z_k - h(X_{k|k-1})\right)$ Eq.\ref{R3}} & \makecell{Cost function \\sometimes oscillate}  \\ \hline

\makecell{MS Method*} & \makecell{Scaled up $P_{N|N}$ (Eq.\ref{P1}) \\ trimmed to [0,0;0,\checkmark] } & \makecell{MS Method (Eq.\ref{Q2})\\ trimmed to [\checkmark, 0; 0, 0]} & \makecell{Filtered residue \\$\left(Z_k - h(X_{k|k})\right)$ Eq.\ref{R2}} & \makecell{\textbf{R} can go to \\ a very high value}  \\ \hline

\multicolumn{5}{c}{  }\\
\multicolumn{5}{c}{*Since in MS and MT methods the $\mathbf{P_0}$ is not specified we have suggested the above}

\end{tabu}}
\end{table}
\end{center}

%%%%%%%%%%%%%%%%%%%%%%%%%%%%%%%%%%%%%%%%%%%%

\begin{landscape}
\begin{table}[h]
\vspace{14pt}
\renewcommand\thetable{2a}
\caption{Sensitivity Study : (\textbf{Q} = 0). No. of iterations=20, No. of simulations=50.}{}
\label{tbsmd}
\begin{center}
\begin{footnotesize}
\begin{tabular}{| c | c | c | c | c | c | c | c | c | c | c |}
\hline
Study &
\makecell{$\Theta$ ratio\\EKF/NR} &
\makecell{CRB ratio\\EKF/NR} &
\makecell{Consistency\\ ratio-EKF}&
\makecell{Consistency\\ ratio-NR}&
\makecell{Spread\\factor\\EKF} &
\makecell{Spread\\factor\\NR} &
\makecell{\textbf{R} ratio\\EKF/NR} &
\makecell{$\mu$\ of \\\textbf{J1-J5}} &
\makecell{$\sigma$ of \\\textbf{J1-J5}}
\\ \hline

\makecell{$\mathbf{P_0}$ : Scaled up-[0,0;0,\checkmark]\\\textbf{Q} : EM-[\checkmark,0;0,0] \\ \textbf{R} : EM-diag}
&\multicolumn{9}{c|}{Reference adaptive EKF used for Q $>$ 0 case gives extremely slow convergence of \textbf{Q} taking hundreds of iterations.}  \\ \specialrule{.2em}{.0em}{.0em}

\multicolumn{1}{!{\vrule width 2pt}c|}{
\makecell{$\mathbf{P_0}$ : Scaled up-[0,0;0,\checkmark]\\\textbf{Q} : $10^{-10}$-[\checkmark,0;0,0] \\ \textbf{R} : EM-diag} } &
\makecell{ 1.0007  \\  1.0000  \\  0.9869} & \makecell{ 1.0048   \\ 0.9832  \\  0.9830} &
\makecell{ 1.0140  \\  1.3764 \\   1.0919} & \makecell{  1.0533 \\   1.3138 \\   1.0817} &
\makecell{0.8128  \\  1.5488 \\  14.5838} & \makecell{0.8126  \\  1.5079 \\  14.5843} &
\makecell{ 1.0139 \\  1.0128 } &
\makecell{1.9704\\    1.9702  \\  1.9999 \\   0.0048 \\ -10.3911}&
\multicolumn{1}{|c!{\vrule width 2pt}}{\makecell{ 0.0512  \\  0.0512  \\  0.0013  \\  0.0008  \\  0.2243  }}
\\ \specialrule{.2em}{.0em}{.0em}

%----------------------------------------------------------

\makecell{$\mathbf{P_0}$ : Scaled up-diag\\\textbf{Q} : $10^{-10}$-[\checkmark,0;0,0] \\ \textbf{R} : EM-diag}  &
\makecell{0.9998  \\  1.0024 \\   1.0255} & \makecell{1.2439  \\  1.3877  \\   1.7525} &
\makecell{1.0518  \\  1.2966  \\  1.1761} & \makecell{ 1.0533 \\   1.3138  \\  1.0817} &
\makecell{ 1.0223  \\  2.1729  \\ 26.6449} & \makecell{0.8126  \\  1.5079  \\ 14.5843}&
\makecell{ 1.0117 \\   1.0061 }  &
\makecell{1.9530 \\   1.9534  \\  1.9998  \\  0.0047 \\ -10.3252 } &
\makecell{0.0570  \\  0.0570 \\   0.0050 \\   0.0008 \\   0.2308 }\\ \hline

\makecell{$\mathbf{P_0}$ : Scaled up-full\\\textbf{Q} : $10^{-10}$-[\checkmark,0;0,0] \\ \textbf{R} : EM-diag}  &
\makecell{ 0.9991  \\  1.0023 \\   1.0471} & \makecell{ 0.9487 \\   1.1580  \\  1.0710} &
\makecell{ 1.3638 \\   1.5696 \\   1.8936} & \makecell{1.0533   \\ 1.3138  \\  1.0817} &
\makecell{0.8922 \\   2.0313 \\  21.6826} & \makecell{ 0.8126 \\   1.5079 \\  14.5843} &
\makecell{  1.0045  \\  1.0006 } &
\makecell{1.9668  \\  1.9670  \\  1.9999  \\  0.0047  \\ -10.4051   }&
\makecell{0.0438 \\   0.0438  \\  0.0014 \\   0.0008 \\   0.2293  }\\ \hline

\makecell{$\mathbf{P_0}$ : IIM-[0,0;0,\checkmark]\\\textbf{Q} : $10^{-10}$-[\checkmark,0;0,0] \\ \textbf{R} : EM-diag}  &
\makecell{ 1.0014  \\  0.9990  \\  0.9667} & \makecell{  1.0066  \\  0.9942  \\  0.9792} &
\makecell{ 1.1296  \\  2.2338   \\ 1.3654} & \makecell{1.0533  \\  1.3138 \\   1.0817} &
\makecell{0.8997  \\  2.3360 \\  17.3808} & \makecell{0.8126  \\  1.5079 \\  14.5843}&
\makecell{ 1.0228 \\   1.0204 } &
\makecell{  1.9690 \\   1.9710   \\ 1.9998 \\   0.0050  \\-10.2078 }&
\makecell{0.0776  \\  0.0776 \\   0.0015 \\   0.0009 \\   0.2349 } \\ \hline

\makecell{$\mathbf{P_0}$ : IIM-diag\\\textbf{Q} : $10^{-10}$-[\checkmark,0;0,0] \\ \textbf{R} : EM-diag}  &
\makecell{ 0.9995  \\  1.0023 \\   1.0375} & \makecell{1.2489 \\   1.386  \\  1.7504} &
\makecell{ 1.0476  \\  1.5883  \\  1.1881} & \makecell{1.0533  \\  1.3138 \\   1.0817} &
\makecell{  1.0205 \\   2.4209  \\ 26.8883} & \makecell{0.8126  \\  1.5079 \\  14.5843} &
\makecell{  1.0160 \\   1.0102 } &
\makecell{1.9530  \\  1.9542  \\  1.9998 \\   0.0049 \\ -10.1918}&
\makecell{ 0.0655  \\  0.0655  \\  0.0049  \\  0.0009   \\ 0.2403 } \\ \hline

\makecell{$\mathbf{P_0}$ : IIM-full\\\textbf{Q} : $10^{-10}$-[\checkmark,0;0,0] \\ \textbf{R} : EM-diag}  &
\makecell{0.9997  \\  1.0025  \\  1.0340} & \makecell{1.2493  \\ 1.7885 \\   1.7501} &
\makecell{1.0479   \\ 1.6662  \\  1.1851} & \makecell{1.0533  \\  1.3138 \\   1.0817} &
\makecell{1.0185 \\   2.4988 \\  26.7966} & \makecell{0.8126  \\  1.5079 \\  14.5843} &
\makecell{ 1.0178 \\   1.0116 } &
\makecell{1.9540  \\  1.9532 \\   1.9999 \\   0.0050 \\ -10.2001 }&
\makecell{ 0.0656   \\ 0.0656  \\  0.0050  \\  0.0009  \\  0.2420 } \\ \hline

\end{tabular}
\end{footnotesize}
\end{center}
\end{table}
\end{landscape}

%\addtocounter{table}{-1}
\begin{landscape}
\begin{table}[h]
\renewcommand\thetable{2b}
\caption{Sensitivity Study : (\textbf{Q} = 0). No. of iterations=20, No. of simulations=50.}{}
\label{tbsmdb}
\begin{center}
\begin{footnotesize}
\begin{tabular}{|c| c| c| c| c| c| c|c|c|c|c|}
\hline
Study &
\makecell{$\Theta$ ratio\\EKF/NR} &
\makecell{CRB ratio\\EKF/NR} &
\makecell{Consistency\\ ratio-EKF}&
\makecell{Consistency\\ ratio-NR}&
\makecell{Spread\\factor\\EKF} &
\makecell{Spread\\factor\\NR} &
\makecell{\textbf{R} ratio\\EKF/NR} &
\makecell{$\mu$\ of \\\textbf{J1-J5}} &
\makecell{$\sigma$ of \\\textbf{J1-J5}}
\\ \hline

%----------------------------------------------

\makecell{$\mathbf{P_0}$ : Smoothed-[0,0;0,\checkmark]\\\textbf{Q} : $10^{-10}$-[\checkmark,0;0,0] \\ \textbf{R} : EM-diag}  &
\makecell{ 1.0006  \\  1.0002  \\  0.9911} & \makecell{0.2361 \\  0.2306   \\ 0.2315} &
\makecell{ 4.2997  \\  5.7348  \\  4.5352} & \makecell{1.0533  \\  1.3138 \\   1.0817} &
\makecell{  0.5240  \\  1.0375  \\  9.5986} & \makecell{0.8126  \\  1.5079 \\  14.5843} &
\makecell{0.9990  \\  0.9997 } &
\makecell{ 1.9982  \\  1.9982  \\  1.9999  \\  0.0048 \\ -10.5276  }&
\makecell{ 0.0009 \\   0.0009  \\  0.0001  \\  0.0008  \\  0.2240 } \\ \hline

\makecell{$\mathbf{P_0}$ : Smoothed-diag\\\textbf{Q} : $10^{-10}$-[\checkmark,0;0,0] \\ \textbf{R} : EM-diag}  &
\makecell{1.0021 \\   0.9999  \\  0.9560} & \makecell{0.0991 \\   0.2380 \\  0.0993} &
\makecell{10.8182  \\  7.4621 \\  16.1755 } & \makecell{1.0533  \\  1.3138 \\   1.0817} &
\makecell{0.5565 \\   1.5098 \\  14.2577} & \makecell{0.8126  \\  1.5079 \\  14.5843} &
\makecell{ 0.9904  \\  0.9872 } &
\makecell{ 1.9960 \\   1.9960  \\  1.9997  \\  0.0047 \\ -10.5493  }&
\makecell{ 0.0016  \\  0.0016 \\   0.0002  \\  0.0008  \\  0.2294   } \\ \hline

\makecell{$\mathbf{P_0}$ : Smoothed-full\\\textbf{Q} : $10^{-10}$-[\checkmark,0;0,0] \\ \textbf{R} : EM-diag}  &
\makecell{  0.9997  \\  1.0011 \\   1.0260} & \makecell{0.3089  \\  0.3272 \\   0.4565} &
\makecell{4.1863 \\   5.4617 \\   4.3984} & \makecell{1.0533  \\  1.3138 \\   1.0817} &
\makecell{  0.6813 \\   1.5528 \\  18.0659} & \makecell{0.8126  \\  1.5079 \\  14.5843} &
\makecell{  0.9891  \\  0.9859 } &
\makecell{1.9970\\    1.9971 \\   1.9998  \\  0.0047  \\-10.5523  }&
\makecell{0.0011  \\  0.0011 \\   0.0003 \\   0.0008  \\  0.2290   } \\ \hline

\end{tabular}
\end{footnotesize}
\end{center}

\vspace{1cm}
\renewcommand\thetable{3}
\caption{Sensitivity Study : (\textbf{Q} $>$ 0). No. of iterations=100, No. of simulations=50.}{}
\label{tbsmdQ}
\begin{center}
\begin{footnotesize}
\begin{tabular}{|c| c| c| c| c| c| c|c|c|}
\hline
Study &
\makecell{$\Theta$ ratio\\EKF/True} &
\makecell{Consistency\\ ratio-EKF}&
\makecell{Spread\\factor\\EKF} &
\makecell{\textbf{R} ratio\\EKF/True} &
\makecell{\textbf{Q} ratio\\EKF/True}  &
\makecell{$\mu$\ of \\\textbf{J1-J8}} &
\makecell{$\sigma$ of \\\textbf{J1-J8}}
\\ \hline

\makecell{$\mathbf{P_0}$ : Smoothed-full\\\textbf{Q} : EM-full \\\textbf{R} : EM-full}  & \multicolumn{7}{c|}{\makecell{Extended EM algorithm (Bavdekar et al. (2011) : Cost functions diverges after few iterations.\\There is a need for a precise stopping condition without which unity ratios cannot be achieved.}}  \\ \specialrule{.2em}{.0em}{.0em}

\multicolumn{1}{!{\vrule width 2pt}c|}{
\makecell{$\mathbf{P_0}$ : Scaled up-[0,0;0,\checkmark]\\\textbf{Q} : EM-[\checkmark,0;0,0] \\\textbf{R} : EM-diag}}  &
\makecell{ 0.9933 \\   1.0201 \\   1.0764} &
\makecell{1.0327  \\  1.0772  \\  0.9128} & \makecell{7.9434 \\  23.5655 \\  98.4334} &
\makecell{  0.9450  \\  0.9135} &
\makecell{ 1.0648 \\   1.0922} & \makecell{  1.9650\\1.9700\\1.9982\\ 0.0709\\-8.7886\\1.9439\\1.9533\\1.9585} & \makecell{ 0.0224 \\   0.0217\\    0.0091 \\  0.0396 \\  0.2281 \\ 0.0422 \\   0.0618 \\   0.0373}
%\multicolumn{1}{|c!{\vrule width 2pt}}{\makecell{Ratios are close to unity.\\Cost functions converge to\\ their expected values.}}

 \\ \specialrule{.2em}{.0em}{.0em}

\makecell{$\mathbf{P_0}$ : Smoothed-diag\\\textbf{Q} : EM-diag \\\textbf{R} : EM-diag}   &
\makecell{  1.0210  \\  1.0018 \\   0.6806} &
\makecell{ 13.3560 \\  10.0828 \\  13.6464} & \makecell{ 3.9686 \\  13.6349  \\ 51.6473} &
\makecell{ 0.9502 \\   0.9703} & \makecell{1.0286 \\   0.8676} &
\makecell{ 1.9988\\ 1.9996\\ 1.9998\\0.0656\\-8.9374 \\ 2.0031\\ 2.0027\\1.9976} & \makecell{0.0241\\0.0241\\0.0076  \\  0.0325 \\   0.2290  \\  0.0147   \\ 0.0239\\ 0.0282 }

%& \makecell{No stoping condition.\\ The low spread factor is deceptive\\since there is no consistency in the\\ parameter estimates with its covariance.}

\\ \hline

\end{tabular}
\end{footnotesize}
\end{center}
\end{table}
\end{landscape}

\newpage
%%%%%%%%%%%%%%%%%%%%%%%%%%%%%%%%%%%%%%%%%%%%%
%\subsection{SMD System Figures (\textbf{Q} $>$ 0) }

\begin{figure}[h]
\includegraphics[width=6in,height=2in]{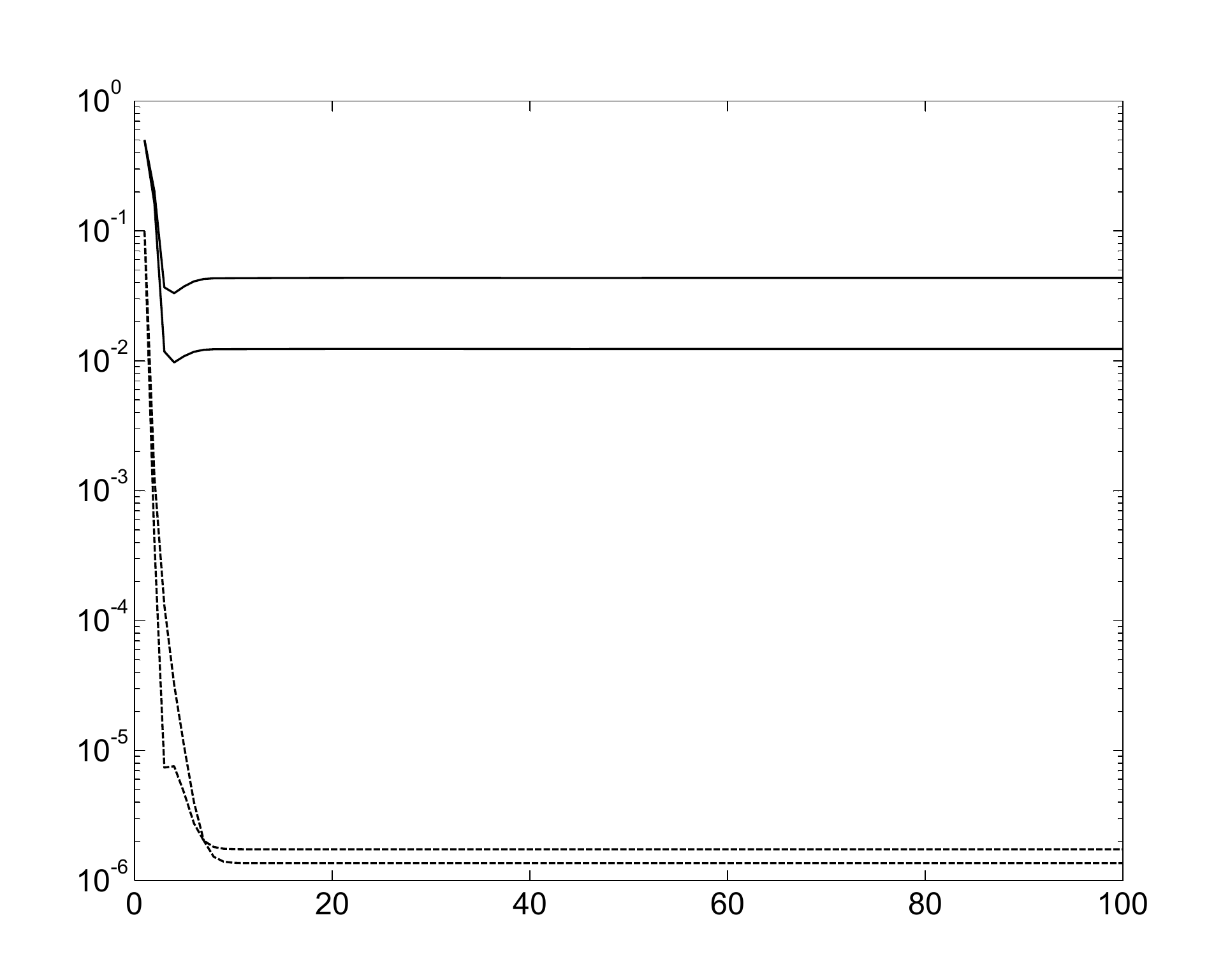}
\caption{Variation of \textbf{Q} (dashed) and \textbf{R} (continuous) with iterations using MS method}
\label{MS_SMD_Q}
\end{figure}

\begin{figure}[h]
\includegraphics[width=6in,height=2in]{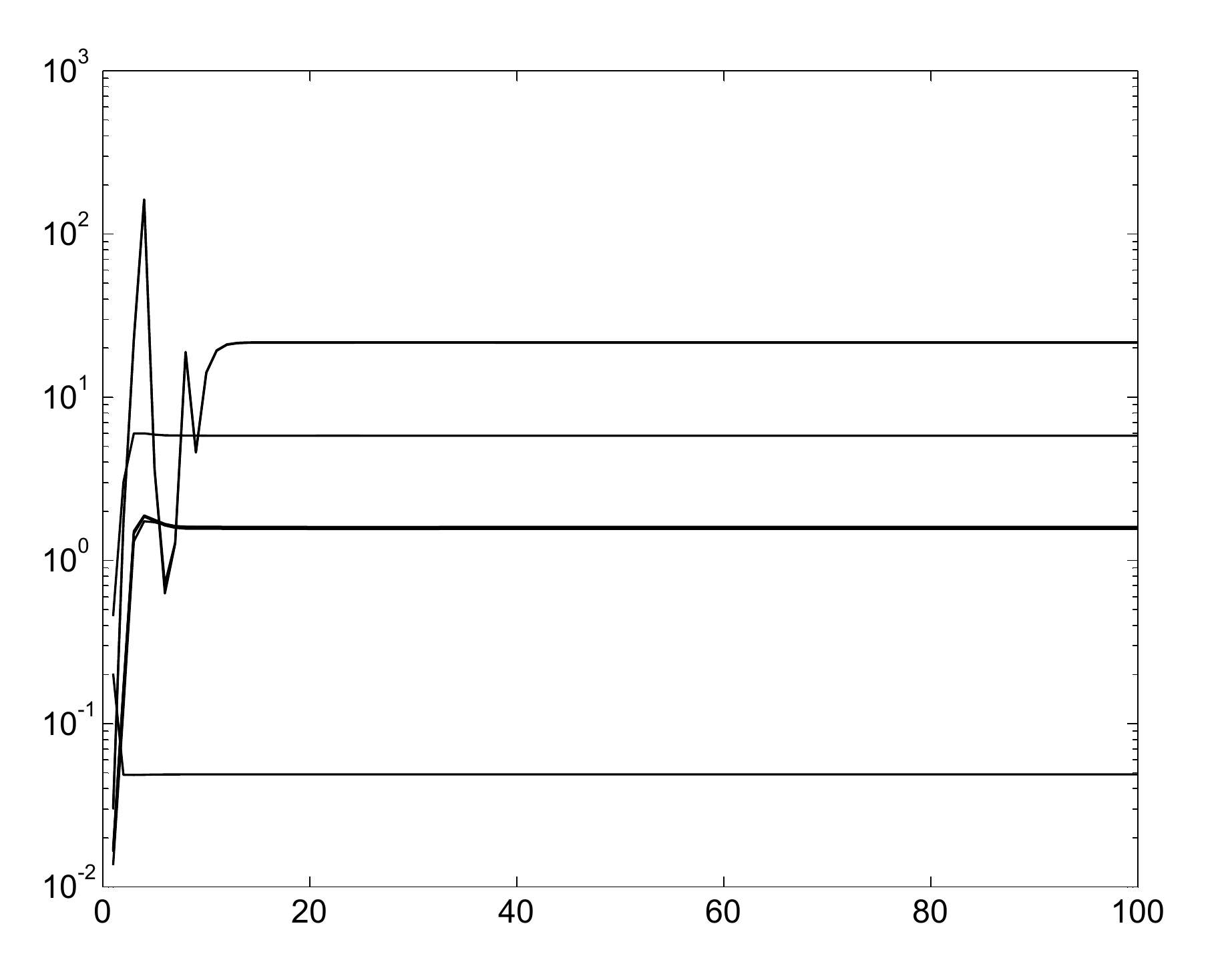}
\caption{Variation of different costs(\textbf{J1-J8}) with iterations using MS method}
\label{MS_SMD_J}
\end{figure}

\begin{figure}[h]
\includegraphics[width=6in,height=2in]{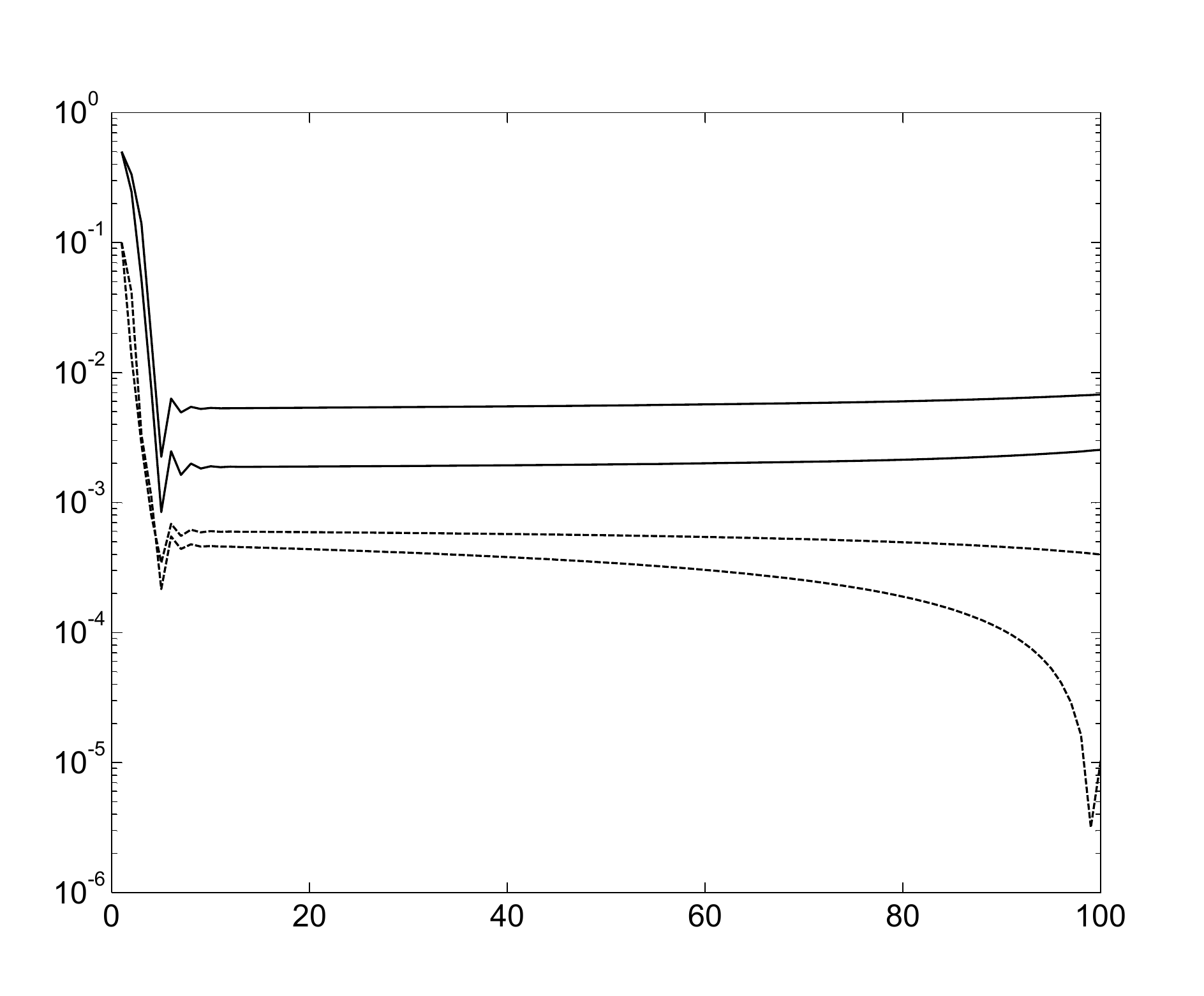}
\caption{Variation of \textbf{Q} (dashed) and \textbf{R} (continuous) with iterations using MT method}
\label{MT_SMD_Q}
\end{figure}

\begin{figure}[h]
\includegraphics[width=6in,height=2in]{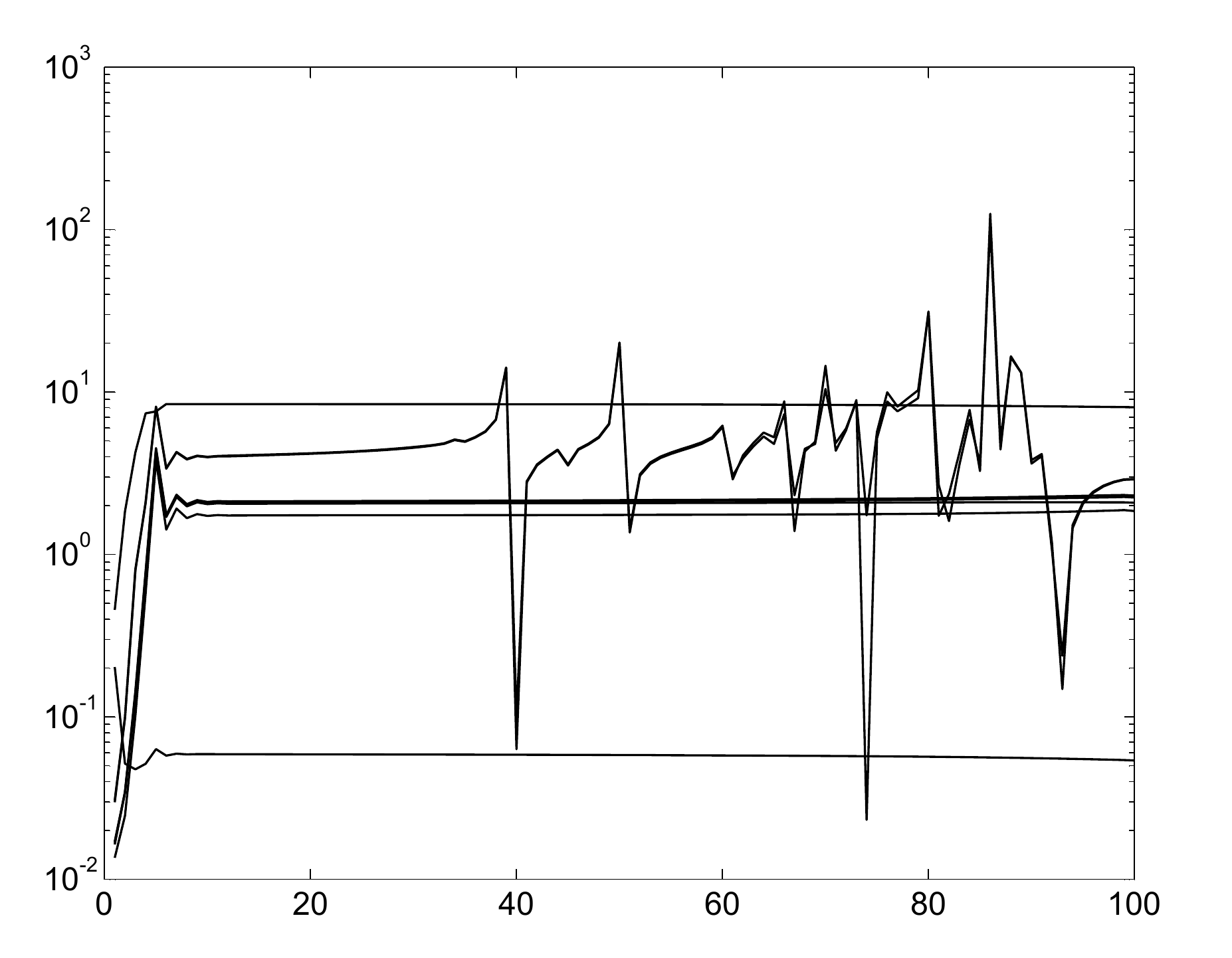}
\caption{Variation of different costs(\textbf{J1-J8}) with iterations using MT method}
\label{MT_SMD_J}
\end{figure}

\begin{figure}[h]
\includegraphics[width=6in,height=2in]{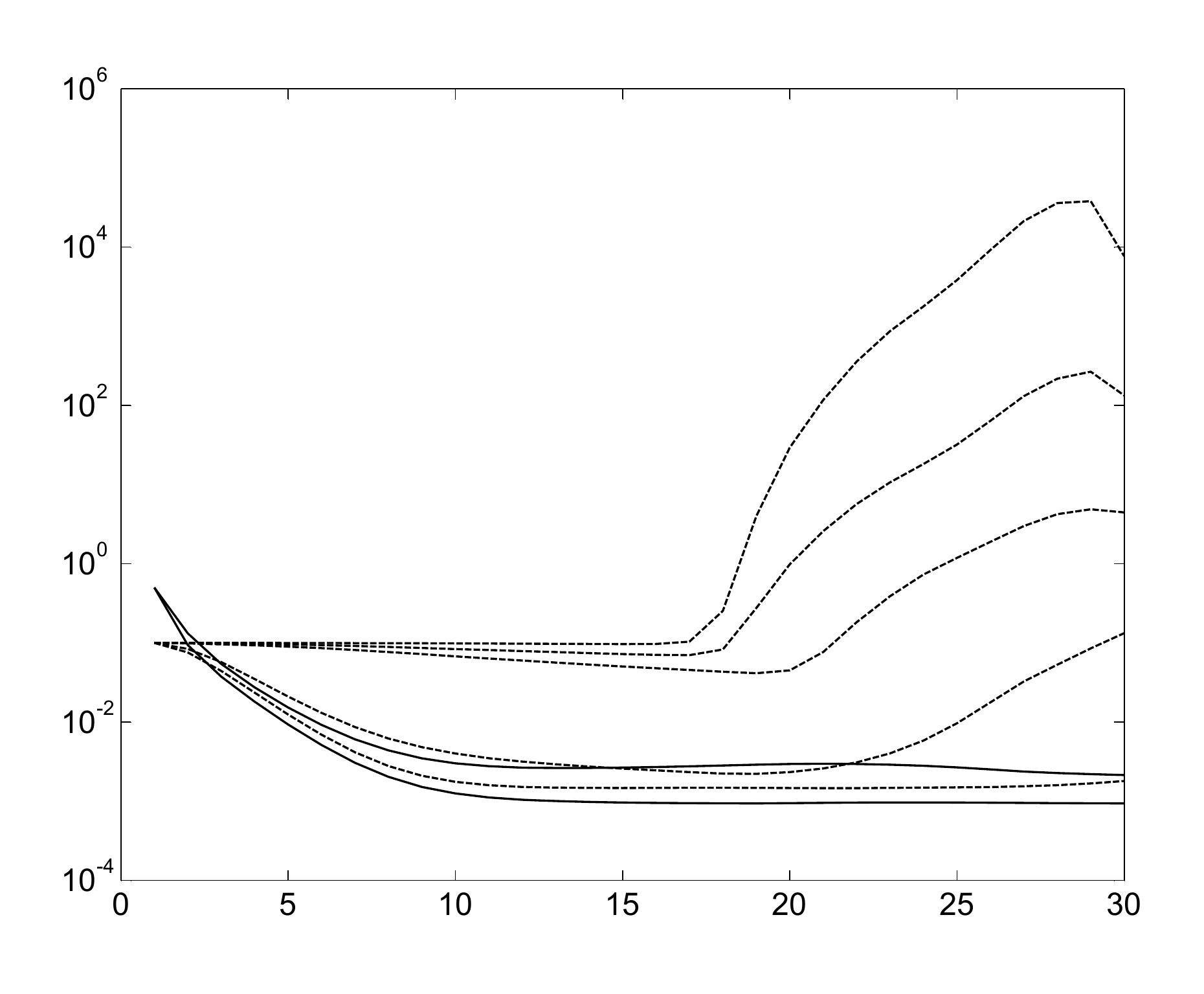}
\caption{{\small Variation of \textbf{Q} (dashed) and \textbf{R} (continuous) with iterations using Bavdekar et al. method}}
\label{EM_SMD_Q}
\end{figure}

\begin{figure}[h]
\includegraphics[width=6in,height=2in]{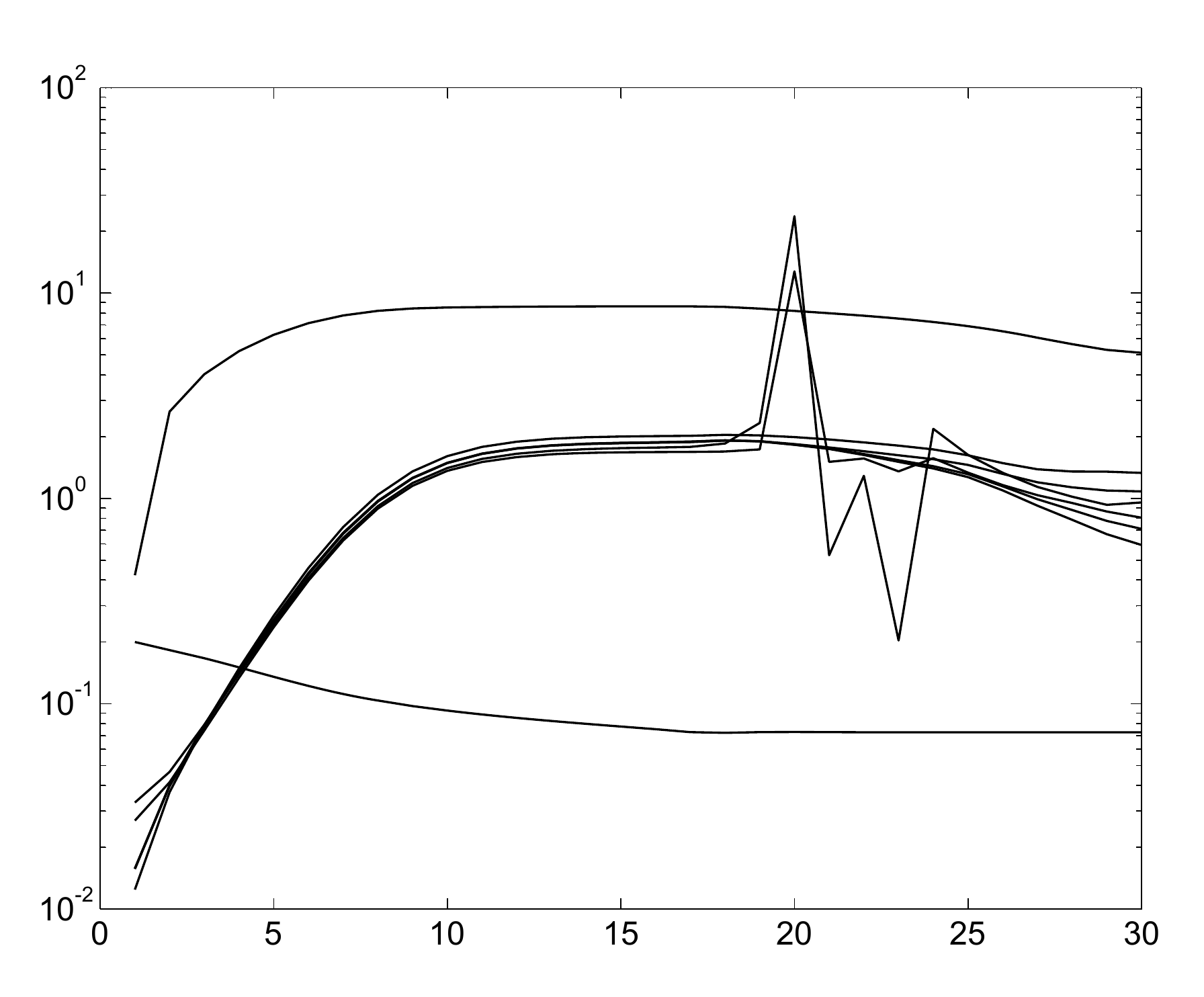}
\caption{{\small Variation of different costs(\textbf{J1-J8}) with iterations using Bavdekar et al. method}}
\label{EM_SMD_J}
\end{figure}

\clearpage
\section{Conclusions}
\label{conclusions}

A comparative study between the existing adaptive techniques suggested by Myers and Tapley, Mohamed and Schwarz and Bavdekar et al. is carried out and a reference recursive recipe (RRR) for tuning the Kalman filter is proposed. A new statistic for the estimation of \textbf{Q} based upon the difference between the stochastic and dynamic trajectories (DSDT) was introduced based on extended EM method. The different cost functions (\textbf{J1-J8}) helps the user to move away from deceptive to decisive convergence. The proposed RRR achieves the Cramer Rao Bound (CRB) of the unknown parameters and provide statistically equilibrium solution for the adaptive filtering problem in an easy and simple way without any need for direct optimization in a few filter iterations through the data. In the menacing problem of tuning the filter statistics, the importance of $\mathbf{P_0}$ has been demonstrated and the notorious \textbf{Q} has been successfully handled. This RRR is believed to be sufficiently simple and general to be used in any Kalman filtering problem.

\section*{Acknowledgments}
Our grateful thanks are due to Profs. R. M. Vasu (Dept. of Instrumentation and Applied Physics), D. Roy (Dept. of Civil Engineering), and M. R. Muralidharan (Supercomputer Education and Research Centre) for help in a number of ways without which this work would just not have been possible at all and also for providing computational facilities at the IISc, Bangalore.

\begin{appendices}
\section{List of Symbols$^*$}
\begin{table}[h]
\begin{center}
\begin{tabular}{ l  l  }
$x$ & State vector of size $n\times 1$\\
$\Theta$ & Parameter vector of size $p\times 1$\\
$X=[x,\Theta]^T$ & Augmented state vector of size $(n+p)\times 1$\\
$Z_k$ & Measurement Vector of size $m\times 1$ at discrete time index `k'\\
$\mathbf{X_0,P_{0}}$ & Initial state and its covariance\\
\textbf{R, Q} & Measurement and Process noise covariance matrix\\
\textbf{J0-J8} & Cost functions\\
$X_{k|k-1}$ & Prior state estimate at time index k based on data upto k-1\\
$X_{k|k}$ & Posterior state estimate at time index k based on data upto k \\
$X_{k|N}$ & Smoothed state estimate at time index k based on data upto N\\
$X{d_{k|N}}$ & Dynamical state estimate at time index k based on data upto N\\
$P_{k|k-1}$ & Prior state covariance matrix at time index k given data upto k-1\\
$P_{k|k}$ & Posterior state covariance matrix at time index k given data upto k\\
$P_{k|N}$ & Smoothed state covariance matrix at time index k given data upto N\\
$P_{k,k-1|N}$ & Lag one covariance of the smoothed state estimate\\
$K_k$ & Kalman Gain based on data upto time index k\\
$K_{k|N}$ & Smoothed Gain based on all data upto time index N\\
$F_{k-1}=\left[\frac{\partial{f}}{\partial{X}}\right]_{X=X_{k-1|k-1}}$ & State Jacobian matrix using posterior state estimate\\
$F_{k-1|N}=\left[\frac{\partial{f}}{\partial{X}}\right]_{X=X_{k-1|N}}$ & State Jacobian matrix using smoothed state estimate\\
$F{d_{k-1|N}}=\left[\frac{\partial{f}}{\partial{X}}\right]_{X=X{d_{k-1|N}}}$ & State Jacobian matrix using dynamical state estimate\\
$H_{k}= \left[\frac{\partial{h}}{\partial{X}}\right]_{X=X_{k|k-1}}$ & Measurement Jacobian matrix using prior state estimate\\
$H_{k|k}= \left[\frac{\partial{h}}{\partial{X}}\right]_{X=X_{k|k}}$ & Measurement Jacobian matrix using posterior state estimate\\
$H_{k|N}= \left[\frac{\partial{h}}{\partial{X}}\right]_{X=X_{k|N}}$ & Measurement Jacobian matrix using smoothed state estimate\\
\end{tabular}
\begin{flushleft}
\textit{*Most other symbols are explained as and when they occur. }
%\end{center}
\end{flushleft}
\end{center}
\end{table}
\end{appendices}

\clearpage

\begin{footnotesize}
%\section*{References}

\begin{itemize}

\item[] \textbf{{\large References}}

\item[] Alspach, D. (1974) A parallel filtering algorithm for linear systems with unknown time varying noise statistics. \emph{IEEE Trans. Auto. Cont., 19(5): 552-556}.

\item[] Ananthasayanam, M. R.,  Suresh, H. S. and  Muralidharan, M. R. (2001) GUI based software for teaching parameter estimation technique using MMLE. \emph{Report 2001 FM 1,Project-JATP/MRA/053,IISc-Bangalore}.

\item[] Ananthasayanam,  M. R.,  Anilkumar, A. K. and  Subba Rao, P. V. (2006) New Approach for the Evolution and Expansion of Space Debris Scenario. \emph{Journal of Spacecraft and Rockets, Vol. 43, No. 6 : pp. 1271-1282}.

\item[]  Ananthasayanam, M. R. and  Bharadwaj, K. M. (2011) The Philosophy, Principles, and Practice of Kalman Filter since Ancient Times to the Present, " \emph{45th IAA History of Astronautics Symposium}.

\item[] Anilkumar, A. K. (2000) Application of Controlled Random Search Optimisation Technique in MMLE with Process Noise. \emph{MSc Thesis, Dept. of Aerospace Engineering IISc, Bangalore}.

\item[] Bar-Shalom, Y.  and  Li, X. (1993)  Estimation and Tracking : Principles, Techniques and Software. \emph{Boston : Artech House Radar Library}.

\item[]  Bar-Shalom, Y.,  Rong Li, X. and  Kirubarajan, T. (2001) Estimation with Applications To Tracking and Navigation, Theory, Algorithm and Software.  \emph{John Wiley and Sons. Inc}.

\item[] Bavdekar, V. A.,  Deshpande, A. P. and  Patwardhan, S. C. (2011) Identification of process and measurement noise covariance for state and parameter estimation using extended Kalman filter. \emph{Journal of Process control, 21 : 585-601}.

\item[]  Belanger, P. R. (1974) Estimation of noise covariances for a linear time-varying stochastic process. \emph{Automatica Volume 10, Issue 3, May 1974, Pages 267-275}.

\item[]  Bohlin, T. (1976) Four cases of identification of changing systems. \emph{System Identification: Advances and Case Studies, Academic Press, 1st edition}.

\item[]  Bohn, C. (2000) Recursive Parameter Estimation for Non Linear Continuous time systems through Sensitivity-Model based Adaptive Filters. \emph{PhD Thesis, Dept. of Electrical Engineering and Information Science}.

\item[] Brown, R. and  Hwang, P. (2012) Introduction to Random Signals and Applied Kalman Filtering, With MATLAB Exercises, 4$^{th}$ Edition. \emph{John Wiley and Sons, Inc.}

\item[] Candy, J.V. (1986) Signal Processing The Model based approach. \emph{Mcgraw Hill Series in Electrical and Computer Engineering}.

\item[] Carew, B. and Belanger, P. R. (1973) Identification of optimum filter steady state gain for systems with unknown noise covariances. \emph{IEEE Trans. Auto. Cont., 18(6): 582-587}.

\item[] Costagli, M. and  Kuruoglu, E. E. (2007) Image separation using particle filters. \emph{Digital Signal Processing 17, pp. 935-946}.

\item[] Datta, G.S. and  Mukerjee, R. (2005) Probability Matching Priors: Higher Order Asymptotics. \emph{Series: Lecture Notes in Statistics, Vol. 178}.

\item[] Evensen, G. (2009) Data Assimilation: The Ensemble Kalman Filter, Second Edition. \emph{Springer Verlag }.

\item[]  Federer, W. T. and  Murthy. B. R. (1998) Kalman Filter Bibliography :  Agriculture, Biology, and Medicine. \emph{Technical Report BU-1436-M Department of Biometrics, Cornell University, Ithaca}.

\item[]  Fruhwirth, R., Regier, M., Bock, R. K., Grote H. and Notz, D. (2000) Data Analysis  Techniques for High-Energy Physics. \emph{Cambridge Monographs on Particle Physics, Nuclear physics and Cosmology}.

\item[] Gemson, R. M. 0. (1991) Estimation Of Aircraft Aerodynamic Derivatives Accounting For Measurement And Process Noise By EKF Through Adaptive Filter Tuning. \emph{PhD Thesis, Dept. of Aerospace Engineering IISc, Bangalore}.

\item[]  Gemson, R. M. O. and  Ananthasayanam, M. R. (1998) Importance of Initial State Covariance Matrix for the Parameter Estimation Using Adaptive Extended Kalman Filter. \emph{AIAA-98-4153, pp. 94-104}.

\item[] Grewal, M. S., Lawrence, R.W. and Andrews, A. P., (2007) Global Positioning Systems, Inertial Navigation, and Integration, Second Edition. \emph{John Wiley \& Sons, Inc., Publication}.

\item[] Hilborn, C. and Lainiotis, D. (1969) Optimal estimation in the presence of unknown parameters. \emph{IEEE Trans. Systems, Science, and  Cybernetics, 5(1):38-43}.

\item[] Julier, S.J., Uhlmann, J.K., Durrant-Whyte, H.F. (1995) A New Approach for Filtering Nonlinear Systems. \emph{Proceedings of the 1995 American Control Conference, vol.3, pp. 1628- 1632 }.

\item[] Kailath, T. (1970) An Innovation Approach To Detection and Estimation Theory. \emph{Proceedings of the lEEE,  Vol. 58, Issue: 5, pp. 680 -695}.

\item[] Kalman, R. E. (1960) A New Approach to linear Filtering and Prediction Problems. \emph{Transactions of the ASME-Journal of Basic Engineering, 82 (Series D) : pp. 35-45}.

\item[] Kalman, R. E., Bucy, R. S. (1961)  New Results in Linear Filtering and Prediction Theory. \emph{J. Fluids Eng. 83(1), 95-108}.

\item[]  Kashyap, R. (1970) Maximum likelihood identification of stochastic linear systems.\emph{IEEE Trans. Auto. Cont., 15(1), pp. 25-34}.

\item[] Klein, V.  and  Morelli, E.A. (2006)  Aircraft System Identification. Theory and Practice. \emph{AIAA Edu. Series}.

\item[] Kleusberg, A. and  Teunissen, P. J. G. (1996) GPS for Geodesy : First Edition. \emph{Springer Verlag }.

\item[] Lau, T. and  Lin, K. (2011) Evolutionary tuning of sigma-point Kalman filters. \emph{IEEE International Conference on Robotics and Automation (ICRA), pp. 771-776 }.

\item[] Ljung, L. (1979) Asymptotic behaviour of the EKF as a parameter estimator for linear systems. \emph{IEEE trans. Automatic control, Vol. AC 24, pp. 36-50}.

\item[]  Ljungquist, D. and  Balchen, J. G. (1994) Recursive prediction error methods for online estimation in nonlinear state space models. \emph{Modeling, Identification and Control, Vol. 15 No. 2 : pp. 109-121}.

\item[] Maine, R. E. and Iliff, K. W. (1981) Programmer's manual for MMLE3, a general Fortran program for Maximum Likelihood parameter estimation", \emph{NASA TP-1690}.

\item[] Manika, S., Bhaswati, G. and Ratna, G. (2014) Robustness and Sensitivity Metrics for Tuning the Extended Kalman Filter. \emph{IEEE Trans. on Instrumentation and Measurement, Vol. 63, No. 4, pp. 964-971}.

\item[] Maybeck, P. S. (1979) Stochastic Models, Estimation, and Control: Volume 1, " \emph{New York: Academic Press}.

\item[] Mazor,  E., Averbuch,  A., Bar-Shalom, Y.,  Dayan, J. (1998) Interacting Multiple Model methods in Target Tracking: A Survey. \emph{IEEE Transactions on Aerospace and Electronic Systems, Vol. 34 : pp. 103 - 123}.

\item[] Mehra, R. K.  (1970) On the identification of variances and adaptive Kalman filtering. \emph{IEEE Trans. Automat. Control, Vol. 15, N.2, pp. 175-184}.

\item[] Mehra, R. (1972) Approaches to adaptive filtering. \emph{IEEE Trans. Auto. Cont., 17, pp. 903-908}.

\item[]  Mehrotra, K. Mahapatra, P. R. (1997) A Jerk Model for Tracking Highly Maneuvering Targets. \emph{IEEE Transactions on Aerospace and Electronic Systems, VOL. 33, NO. 4}.

\item[]  Mohamed, A. H. and Schwarz,  K. P. (1999)  Adaptive Kalman Filtering for INS/GPS. \emph{Journal of Geodesy,  Volume 73, Issue 4, pp 193-203}.

\item[]  Myers, K. A. and  Tapley, B. D. (1976) Adaptive Sequential Estimation with Unknown Noise Statistics. \emph{IEEE Transactions on Automatic Control, Vol. AC 21 : pp. 520-525}.

\item[]  Neethling, C. and Young, P. (1974) Comments on Identification of optimum filter steady state gain for systems with unknown noise covariances. \emph{IEEE Trans. Auto. Cont., 19(5):623-625}.

\item[]  Odelson, B. J., Lutz, A. and Rawlings, J. B. (2006) The autocovariance-least squares method for estimating covariances: application to model-based control of chemical reactors. \emph{IEEE Trans. Control Syst. Technol. Vol. 14, No. 3, pp. 532-540}.

\item[] Oshman, Y. and  Shaviv, I. (2000) Optimal Tuning of a Kalman Filter Using Genetic Algorithm. \emph{AIAA Paper 2000-4558}.

\item[] Oussalah M., De Schutter J. (2000) Adaptive Kalman Filter for Noise Identification. \emph{International Conference on Noise and Vibration Engineering, ISMA25 : pp. 1225-1232}.

\item[]   Powell, T. D. (2002) Automated tuning of an Extended Kalman Filter Using the Downhill Simplex Algorithm. \emph{J. Guidance, Control,and Dynamics, Vol. 25, No. 5, pp. 901-908}.

\item[] Rauch,  H. E., Tung, F. and Striebel, C. T. (1965)  Maximum Likelihood Estimates of Linear Dynamic Systems. \emph{AIAA Journal, Vol. 3, No. 8, pp. 1445-1450}.

\item[] Shumway, R. H. and Stoffer, D. S. (1982) An approach to time series smoothing and forecasting using the EM algorithm. \emph{J. Time Series Anal., 3, pp. 253-264}.

\item[] Shumway, R. H.,  Stoffer, D. S. (2000) Time Series Analysis and its Applications. \emph{Springer, Verlag, NY}.

\item[] Shyam, M. M. (2014)  An Iterative Tuning Strategy for Achieving Cramer Rao Bound Using Extended Kalman Filter for a Parameter Estimation Problem. \emph{MTech Thesis, Dept. of Electrical Engineering, IIT Kanpur}.

\item[] Shyam, M. M., Naren Naik, Gemson, R. M. O, Ananthasayanam, M. R. (2015) Introduction to the Kalman Filter and Tuning its Statistics for Near Optimal Estimates and Cramer Rao Bound. \emph{TR/EE2015/401, Dept. of Electrical Engineering, IIT Kanpur, \url{http://arxiv.org/abs/1503.04313}}.

\item[] Valappil, J. and Georgakis, C. (2000) Systematic Estimation of State Noise Statistics for Extended Kalman Filters. \emph{AIChe Journal Vol. 46, No. 2,  pp. 292–308}.

\item[]  Visser, H. and  Molenaar, J. (1988) Kalman Filter Analysis in Dendroclimatology. \emph{Biometrics, pp 929-940}.

\item[] Wells, C. (1996) The Kalman Filter in Finance. \emph{Springer-Science+Business Media, BV.}

\item[] Zagrobelny, M. A. and Rawlings, J. B. (2014) Identification of Disturbance Covariances Using Maximum Likelihood Estimation. \emph{TR No. 2014-02, Department of Chemical and Biological Engineering, University of Wisconsin-Madison}.

\end{itemize}

\end{footnotesize}
\end{document}